\documentclass{article}

\usepackage[english]{babel}
\usepackage[a4paper,top=2cm,bottom=2cm,left=2.5cm,right=2.5cm]{geometry}
\usepackage{amsmath,amssymb}
\usepackage{booktabs}
\usepackage{tabularx}
\usepackage{caption}
\usepackage{graphicx}
\usepackage{microtype}
\usepackage[section]{placeins}  % keep floats inside their section
\usepackage[skip=\baselineskip]{parskip}
\usepackage[colorlinks=true,allcolors=blue]{hyperref}
\usepackage{authblk}

\makeatletter
\renewcommand\AB@authnote[1]{\raisebox{0.5ex}{\normalfont\scriptsize #1}}
\renewcommand\AB@affilnote[1]{\raisebox{0.5ex}{\normalfont\scriptsize #1}}
\makeatother

\title{Observer-Based Model Predictive Control for Isoflux Regulation in the EXL-50U Spherical Tokamak}

\author[1]{Chang Lu}
\author[1]{Jiayi Zhi}
\author[1]{Xian Guo}
\author[2,*]{Tianyuan Liu}

\affil[1]{Institute of Robotics and Automatic Information System, Nankai University, Tianjin 300350, China}
\affil[2]{Beijing ENN Fusion Energy Science and Technology Co., Ltd., Beijing 101111, China}
\date{}
\begin{document}
\maketitle
\normalfont
\noindent\textit{* Corresponding author: Tianyuan Liu (liutianyuan@enn.cn).}
\begin{abstract}
Isoflux control in spherical tokamaks requires coordinated regulation of plasma current and boundary shape under limited coil voltages and computational resources. High-dimensional electromagnetic models and unmeasurable internal states complicate millisecond-scale online control. This study develops observer-based constrained model predictive control (MPC) for joint current and boundary regulation in EXL-50U. Vacuum-vessel spatial coarsening reduces a 538-state linearized response model to 40 states while retaining control-relevant electromagnetic responses. A Kalman observer fuses available measurements with model information to provide full-state estimates for MPC, which jointly optimizes current and boundary-flux tracking over a finite horizon with explicit coil-voltage constraints. Matrix precomputation, problem-specific solver code generation, and structured linear-system solution reduce online computational cost.

Closed-loop simulations on a nonlinear free-boundary Grad--Shafranov evolutive model cover nominal operation, measurement noise, cross-configuration model mismatch, a one-step delay, and combined noise and delay, with comparisons against proportional--integral--derivative (PID) control and a linear--quadratic regulator (LQR). Under nominal conditions, MPC achieves a plasma-current root-mean-square (RMS) error of $1.126~\mathrm{kA}$, a last-closed-flux-surface geometric RMS error of $0.0164~\mathrm{m}$, and a maximum channel-wise flux RMS error of $2.463~\mathrm{mWb}$, all below those of the two comparison controllers. Under measurement noise, the boundary error is $0.0199~\mathrm{m}$ over the common evaluation window, and the closed loop sustains regulation throughout the test. The cross-configuration test completes a limiter-to-divertor transition without relinearization or controller retuning, while regulation is also sustained under the one-step delay and combined disturbances. Across the five cases, mean MPC computation times range from $0.413$ to $0.441~\mathrm{ms}$, with all 99th percentiles below $1~\mathrm{ms}$. These results demonstrate joint current and isoflux regulation and millisecond-scale computational feasibility under the tested conditions, providing an algorithmic and implementation basis for subsequent EXL-50U experiments.

\end{abstract}


\section{Introduction}
\label{sec:intro}

Magnetic confinement fusion is a promising route to sustainable low-carbon energy\cite{Magnetic-confinement-fusion,creely2020sparc}. Reliable operation of tokamaks, a major class of magnetic confinement devices, requires real-time regulation of plasma equilibrium and boundary shape\cite{ITER-overview}. Boundary shape affects energy confinement, magnetohydrodynamic stability, divertor heat-load distribution, and the protection of plasma-facing components\cite{Plasma-boundary-phenomena-in-tokamaks,shape-control-review,MHD-control}. Shape control is therefore important for high-current, highly elongated configurations, advanced divertor operation, and long-pulse discharges\cite{Advanced-scenarios}.

EXL-50U offers a flexible poloidal-field coil configuration and an expanding range of spherical-tokamak experimental scenarios\cite{exl-overview,Shi2025EXL50UStrategy,Shi2026EXL50UOverview}, and has achieved mega-ampere hydrogen--boron plasma discharges\cite{Shi2025EXL50U1MA}. Isoflux-control experiments based on real-time equilibrium reconstruction have already been conducted on the device\cite{ptefit}. One previous experiment combined equilibrium reconstruction, feedforward control, and proportional--integral--derivative (PID) feedback, although boundary positioning performance still required improvement.

Boundary regulation also affects the operating conditions of auxiliary heating systems. For example, the coupling resistance of ion cyclotron range of frequencies (ICRF) antennas is sensitive to the antenna--plasma distance and edge density distribution\cite{Hartmann2001ICRFCoupling,Mayoral2007ICRFCoupling,Greene1984ICRF,Gan2018ICRFCoupling}. Local heat fluxes near the antennas also depend on plasma configuration and antenna distance\cite{Jacquet2013ICRFHeatLoads}. These relationships further motivate reproducible boundary control.

Isoflux control translates a desired boundary into feedback targets by regulating the flux at selected boundary control points toward a common boundary-flux reference. Rather than describing the shape through geometric parameters such as elongation and triangularity, this approach represents the desired boundary using control points, providing a flexible interface for different configurations\cite{shape-control-review,east-plasma,DIII-D-shape,JT-shape}. Isoflux control has been applied on devices including TCV and EAST\cite{MIMO_iso,TCV-overview,east-iso}. Expressing the target in terms of flux does not, however, remove plant coupling: the plasma current, poloidal field (PF) circuits, power supplies, and conducting structures jointly determine the boundary response, while coil-voltage limits restrict the available control action\cite{MIMO_iso}. Isoflux control must therefore coordinate plasma current and multiple boundary-flux channels under actuator constraints.

Recent reinforcement-learning studies on EXL-50U address different magnetic-control objectives and validation stages. Xing et al. conducted closed-loop vertical-position experiments and compared the learned controller with the PID controller used in device operation\cite{xing2026reinforcementlearningverticalposition}. Guo et al. trained a control policy using a rigid RZIP state-space model and combined it with real-time position reconstruction for experimental current and position regulation\cite{guo2026ssmrl}. For the more complex X-point target (XPT) configuration, Ding et al. studied coordination of current, boundary, and magnetic-null objectives in an experimentally calibrated free-boundary Grad--Shafranov evolutive (FGE) environment and evaluated policy performance in simulation\cite{ding2026xpt}. Building on this context, the present study investigates model predictive control (MPC) for joint current and boundary-flux regulation through explicit constrained optimization.

MPC provides an optimal-control approach to this constrained multivariable regulation problem. It uses a dynamic model to predict future responses and balances tracking error, control effort, and actuator constraints through receding-horizon optimization\cite{MPC-review,MPC-LQR-PID}. The PID controller and linear--quadratic regulator (LQR) used in this study enforce voltage limits by saturating their control outputs, whereas MPC includes the voltage bounds directly in the optimization problem. This distinction makes constraint handling the principal motivation for adopting MPC.

Previous fusion-control studies support this choice. Work on large fusion devices has demonstrated the feasibility of MPC for plasma-current and magnetic control under actuator constraints\cite{iter-mimo,iter-mpc-shape,demo}. Studies on EAST have further explored predictive coordination of plasma parameters\cite{east-mpc}, while shape-control experiments on TCV have demonstrated boundary tracking and constraint handling with MPC\cite{TCV-iso}. These results support the use of MPC for constrained plasma-current and shape regulation.

Application to a specific device also requires prediction models and state feedback compatible with its computational and measurement resources. For the spherical-tokamak equilibrium-control problem considered here\cite{Spherical-Overview}, high-fidelity models resolving the vessel electromagnetic response contain many internal states, whereas a millisecond control period restricts the budget for state propagation and online optimization. A compact prediction model that retains control-relevant responses is therefore needed. Model reduction alleviates computational demands but does not make all states directly measurable: vessel-induced currents affecting transient responses must still be estimated from available signals to provide a complete initial state for MPC prediction. The closed-loop effects of local linear-model mismatch, measurement noise, and control delay must also be considered. A device-oriented MPC implementation thus requires integration of model reduction, state observation, and online optimization, followed by evaluation under off-nominal conditions.

For the $1~\mathrm{ms}$ EXL-50U control period, this study integrates vacuum-vessel spatial coarsening, Kalman state estimation, and constrained MPC. Nonlinear FGE closed-loop simulations evaluate tracking, applicability under off-nominal conditions, and online computational cost.

\begin{minipage}{\linewidth}
The main contributions are as follows:

(1) An MPC method for EXL-50U isoflux control is developed to explore the application of optimal control to spherical tokamaks.

(2) An output-feedback framework integrating full-state estimation of the reduced model and constrained optimization is established for millisecond-scale control.

(3) Closed-loop simulations under off-nominal conditions assess robustness and feasibility, providing a basis for subsequent EXL-50U experiments.
\end{minipage}

The remainder of this paper is organized as follows. Section 2 introduces the EXL-50U device and technical context, defines the isoflux-control task, and then presents the nonlinear FGE simulator and proposed control framework. Section 3 constructs the compact control model, compares full-order and reduced-order responses, and develops the Kalman observer. Section 4 formulates the constrained MPC problem using estimated states and defines the output-feedback architecture. Section 5 evaluates nominal tracking, controller comparisons, off-nominal responses, and computational performance. Section 6 concludes the paper and outlines future work.

\section{EXL-50U System and Isoflux-Control Problem}

This section introduces the EXL-50U device and control system before defining the joint current and isoflux-control task. It then describes the nonlinear FGE plant and the observer-based MPC framework used to address that task.

\subsection{The EXL-50U Device and Control System}
\label{sec:exl50u_device}

EXL-50U is a compact spherical tokamak used to investigate plasma confinement, equilibrium control, and advanced operating scenarios\cite{exl-overview,Shi2025EXL50UStrategy,Shi2026EXL50UOverview}. It has a flexible poloidal-field coil configuration and has achieved mega-ampere hydrogen--boron plasma discharges\cite{Shi2025EXL50U1MA}.

The device electromagnetic system comprises the central solenoid (CS), PF coils, vertical stabilization (VS) coil, and conducting vacuum vessel. Magnetic fields generated by the coil currents couple to the plasma and jointly determine plasma current and configuration. Changes in plasma and coil magnetic fields also induce vessel currents, making the vessel part of the transient electromagnetic response.

The device control system obtains feedback information from diagnostics, power-supply measurements, and equilibrium reconstruction. Plasma current and coil currents are obtained from diagnostic and power-supply measurements, respectively, while magnetic-axis coordinates and boundary-related fluxes can be supplied by real-time PTEFIT equilibrium reconstruction\cite{ptefit}. Together with the coil power supplies, these signals provide the device interfaces for plasma feedback control.

\begin{figure}[htbp]
\centering
\includegraphics[width=\textwidth]{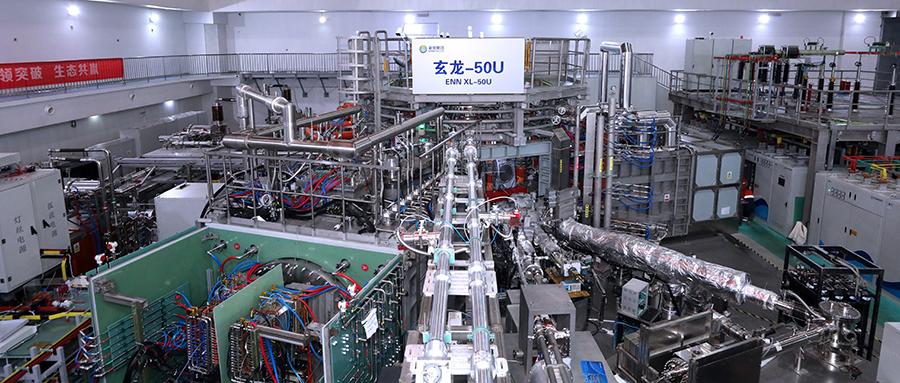}
\caption{The EXL-50U facility.}
\label{fig:overall_technical_framework}
\end{figure}

Figure~\ref{fig:overall_technical_framework} shows the EXL-50U facility, which provides an experimental platform for plasma equilibrium and shape-control research.

EXL-50U has hosted isoflux-control experiments using real-time equilibrium reconstruction\cite{ptefit}, as well as reinforcement-learning-based vertical position control experiments that included comparison with the PID controller used in device operation\cite{xing2026reinforcementlearningverticalposition}.

\FloatBarrier

\subsection{Plasma-Current and Shape-Control Problem}
\label{sec:control_task}

\begin{figure}[ht]
\centering
\includegraphics[width=\textwidth]{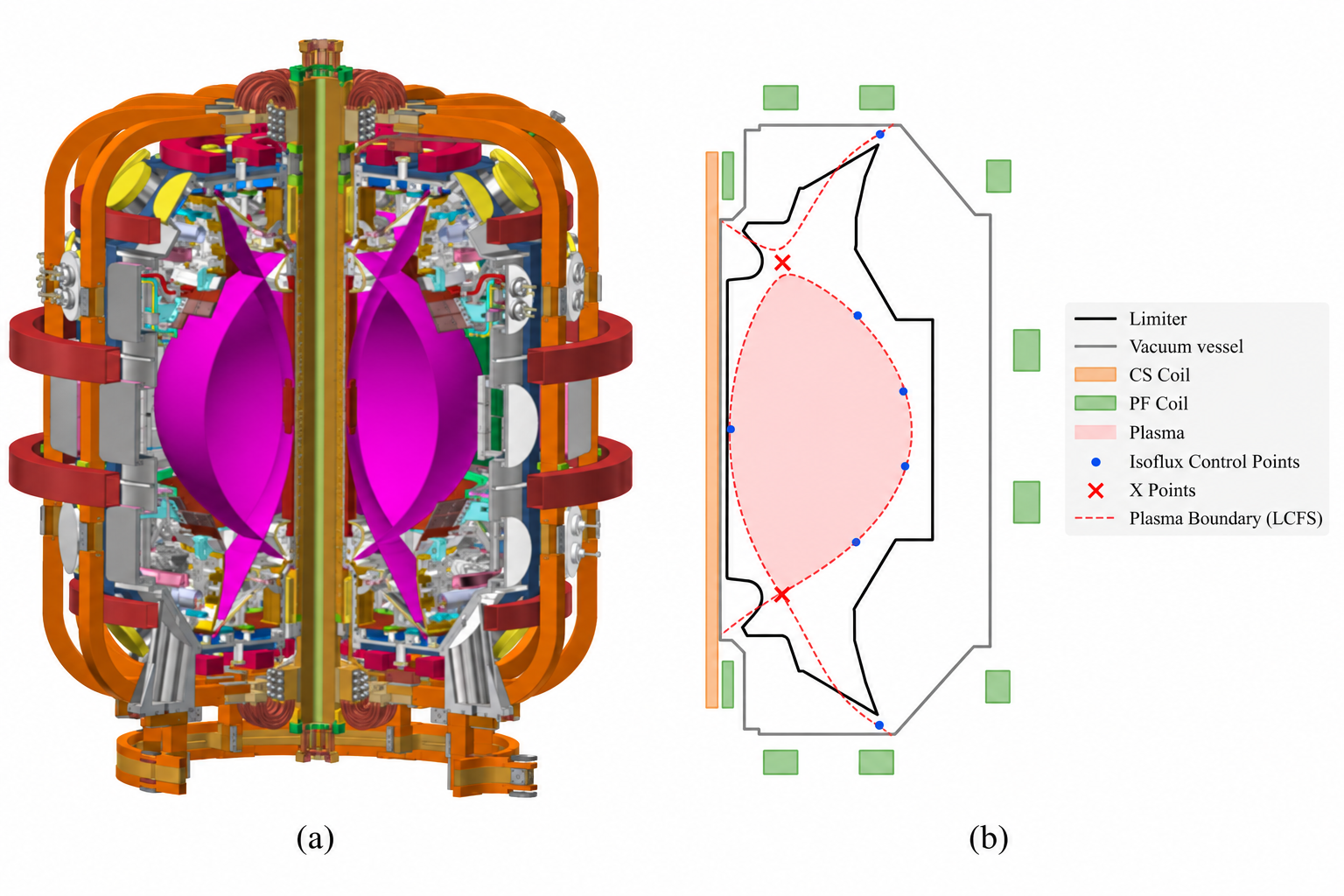}
\caption{EXL-50U device and poloidal cross-section: (a) device structure; (b) coil layout, plasma boundary, and isoflux control points. Red crosses denote X-points.}
\label{fig:exl50u}
\end{figure}

Plasma equilibrium control adjusts the magnetic field through external coils to satisfy prescribed current and boundary-shape requirements. Because coil actuation affects both quantities, they must be coordinated under common actuator constraints. Figure~\ref{fig:exl50u} shows the device structure and poloidal cross-section, including coils, the vessel, boundary control points, and X-points.

The response is governed jointly by plasma equilibrium and external circuit dynamics. Coil voltages drive active-circuit currents, while the plasma and conducting vessel participate through electromagnetic coupling, producing an evolving flux distribution. The voltage-to-current and voltage-to-boundary responses therefore depend on internal electromagnetic states as well as the instantaneous flux at the control points. Denoting these states by $\mathbf{x}$ and external coil voltages by $\mathbf{u}$, the plant is written as

\begin{equation}
\dot{\mathbf{x}}=f(\mathbf{x},\mathbf{u}),
\label{eq:nonlinear_state}
\end{equation}

\begin{equation}
\mathbf{y}=h(\mathbf{x},\mathbf{u}),
\label{eq:nonlinear_output}
\end{equation}
where $f$ describes the coupled state evolution and $h$ maps the states and inputs to the controlled outputs. Here, $\mathbf{y}$ contains plasma current and boundary-control-point fluxes; its explicit form is given in Section 4. The isoflux condition below specifies the shape target and connects it to current regulation.

Under axisymmetry, the poloidal flux field $\psi(R,Z,t)$ describes the plasma configuration, where $R$ and $Z$ are the radial and vertical coordinates. The plasma boundary is the poloidal contour of the last closed flux surface (LCFS). Every point on the actual boundary $\Gamma(t)$ therefore satisfies

\begin{equation}
\psi(R,Z,t)=\psi_{\rm ref}(t),\qquad (R,Z)\in\Gamma(t),
\label{eq:boundary_isoflux}
\end{equation}
where $\psi_{\rm ref}(t)$ is the instantaneous LCFS flux. Isoflux control uses this equal-flux condition to express geometric boundary regulation as flux regulation at desired positions\cite{shape-control-review,MIMO_iso}.

For a desired boundary $\Gamma_{\rm ref}$, the flux at its points should approach the common LCFS reference. Define the residual on the desired boundary as

\begin{equation}
e_{\psi}(R,Z,t)=\psi(R,Z,t)-\psi_{\rm ref}(t),
\qquad (R,Z)\in\Gamma_{\rm ref}.
\label{eq:boundary_flux_residual}
\end{equation}
This residual compares flux at a target position with flux on the actual boundary; it is distinct from the geometric distance between them. The desired contour specifies the shape, whereas $\psi_{\rm ref}(t)$ specifies the common flux level at each instant. Evolution of this flux reference does not move the desired boundary.

A first-order expansion relates the flux residual to local boundary displacement. For a smooth point $\mathbf{q}\in\Gamma(t)$ and a nearby target point $\mathbf{p}\in\Gamma_{\rm ref}$, assuming a locally smooth flux field and $\nabla\psi(\mathbf{q},t)\neq0$,

\begin{equation}
\begin{aligned}
e_{\psi}(\mathbf{p},t)
&=\nabla\psi(\mathbf{q},t)^T(\mathbf{p}-\mathbf{q})
+\mathcal{O}(\|\mathbf{p}-\mathbf{q}\|^2)\\
&\approx\|\nabla\psi(\mathbf{q},t)\|\,\delta n,
\end{aligned}
\label{eq:flux_geometry_relation}
\end{equation}
where $\delta n=\mathbf{n}^T(\mathbf{p}-\mathbf{q})$ is the signed normal displacement from the actual boundary toward the target point, with $\mathbf{n}=\nabla\psi(\mathbf{q},t)/\|\nabla\psi(\mathbf{q},t)\|$. For small displacements and nonzero flux gradients, reducing the residual reflects a reduction in local normal displacement. A given flux error corresponds to different geometric displacements at locations with different gradients. This local relation explains the geometric meaning of isoflux control; it is not a global boundary-coincidence criterion and does not provide a flux-to-distance conversion at X-points, where the gradient vanishes.

To discretize the continuous boundary target, select $N_c$ representative points $\mathbf{p}_i=(R_i^{\rm ref},Z_i^{\rm ref})$ on $\Gamma_{\rm ref}$ and define $\psi_i(t)=\psi(R_i^{\rm ref},Z_i^{\rm ref},t)$. The boundary-regulation errors are

\begin{equation}
e_i(t)=\psi_i(t)-\psi_{\rm ref}(t),\qquad
\boldsymbol{e}_{\psi}(t)=[e_1(t),\ldots,e_{N_c}(t)]^T.
\label{eq:isoflux_error_vector}
\end{equation}
Reducing $\boldsymbol{e}_{\psi}$ drives the target points toward a common boundary-flux contour. Sharing the same reference ensures that the points describe the same boundary target. The finite-point conditions discretize the continuous objective; overall geometric tracking is assessed separately using the LCFS error in Section 5.

The control points remain fixed on the desired contour rather than following the actual boundary. Feedback evaluates the instantaneous flux at these positions against the current LCFS flux. Their number and locations are part of the target-shape discretization.

Current and boundary regulation are physically coupled: external-coil fields, plasma current, and induced vessel currents jointly determine the flux distribution. Changing coil voltages affects both current and boundary responses. Boundary-flux regulation alone therefore cannot replace current tracking, while current regulation alone does not determine the desired shape. For a current reference $I_p^{\rm ref}(t)$, the combined error is

\begin{equation}
\boldsymbol{e}_y(t)=
\begin{bmatrix}
I_p(t)-I_p^{\rm ref}(t)\\
\boldsymbol{e}_{\psi}(t)
\end{bmatrix}.
\label{eq:joint_tracking_error}
\end{equation}
The controller coordinates the reduction of current and flux errors through the same coil voltages, with their relative importance specified in the controller design. Equations~\eqref{eq:nonlinear_state}--\eqref{eq:nonlinear_output} describe the achievable dynamics, and Eq.~\eqref{eq:joint_tracking_error} defines the desired response. Since some internal electromagnetic states are not directly measurable, the feedback design also requires state information inferred from available measurements and the model.

The external-coil voltages are limited by power-supply capabilities:

\begin{equation}
\mathbf{u}_{\min}\leq\mathbf{u}\leq\mathbf{u}_{\max}.
\label{eq:actuator_constraints}
\end{equation}
The transition from the current state to a desired equilibrium is governed by both coupled electromagnetic dynamics and voltage limits, rather than by the static isoflux condition alone. The unified control problem is thus to determine voltages satisfying Eq.~\eqref{eq:actuator_constraints}, using available feedback from the plant in Eqs.~\eqref{eq:nonlinear_state}--\eqref{eq:nonlinear_output}, so as to reduce $\boldsymbol{e}_y$: plasma current tracks its reference while target-point fluxes approach the common LCFS level. Closed-loop responses assess how effectively this objective is achieved.

This formulation retains electromagnetic coupling, output tracking, and actuator constraints without depending on a particular number of control points or controller structure. Section 2.3 describes the nonlinear simulation plant, Section 3 develops the prediction and estimation models, and Section 4 formulates constrained control. Operating settings are specified in Section 5.

\subsection{Nonlinear FGE Simulator}

The nonlinear closed-loop plant is the FGE solver in the Matlab EQuilibrium (MEQ) suite\cite{FGE}. It couples axisymmetric equilibrium, active and passive conductor circuits, and bulk plasma-current evolution to calculate the magnetic response to coil voltages. These three components are described below; a related FGE description is given in Ref.~\cite{ding2026xpt}.

First, the Grad--Shafranov equation determines the poloidal flux distribution. Using the total poloidal flux convention,

\begin{equation}
\Delta^{*}\psi=-2\pi\mu_0Rj_{\phi}
=-4\pi^2\left[\mu_0R^2p'(\psi)+T(\psi)T'(\psi)\right],
\label{eq:fge_grad_shafranov}
\end{equation}
where $\Delta^{*}=R\frac{\partial}{\partial R}\left(\frac{1}{R}\frac{\partial}{\partial R}\right)+\frac{\partial^2}{\partial Z^2}$, $\mu_0$ is the vacuum permeability, $j_\phi$ is the toroidal current density, $p$ is the plasma pressure, $T=RB_\phi$ is the toroidal-field function, and primes denote derivatives with respect to $\psi$. The pressure and toroidal-field profiles determine the current source term. The free-boundary solution determines the flux and LCFS from the plasma-current distribution and external-conductor fields; the desired contour in Section 2.2 is attained through feedback rather than imposed as the actual boundary.

Second, active-coil and passive-conductor currents obey coupled circuit equations. Collecting these currents in $\mathbf{I}_e$ and denoting the discretized plasma-current distribution by $\mathbf{I}_y$ gives

\begin{equation}
\begin{bmatrix}\mathbf{V}_a\\\mathbf{0}\end{bmatrix}
=\mathbf{M}_{ee}\dot{\mathbf{I}}_e
+\mathbf{M}_{ey}\dot{\mathbf{I}}_y
+\mathbf{R}_e\mathbf{I}_e,
\label{eq:fge_conductor_circuits}
\end{equation}
where $\mathbf{V}_a$ contains applied active-circuit voltages, $\mathbf{M}_{ee}$ is the conductor self- and mutual-inductance matrix, $\mathbf{M}_{ey}$ describes conductor--plasma mutual inductance, and $\mathbf{R}_e$ is the conductor resistance matrix. Passive circuits have no applied voltage: their currents arise through induction and decay resistively. These currents modify the equilibrium field, making the vessel response part of the plant dynamics. Here, $\mathbf{V}_a$ denotes the active-circuit interface; Section 4 defines the CS/PF subset optimized by MPC.

The OhmTor-rigid model closes the bulk plasma-current evolution:

\begin{equation}
0=L_p\dot I_p+\mathbf{M}_{pe}\dot{\mathbf{I}}_e
+R_p(I_p-I_{ni}),
\label{eq:fge_plasma_current}
\end{equation}
where $L_p$ is the plasma self-inductance, $\mathbf{M}_{pe}$ is the plasma--conductor mutual inductance, $R_p$ is the effective plasma resistance, and $I_{ni}$ is the non-inductive current component. Together with the spatial current distribution from the equilibrium equation, this relation describes the response of total plasma current to conductor-current changes, inductive effects, and resistive dissipation. Current and boundary flux are consequently outputs of the same coupled system.

At each control step, FGE receives coil voltages, advances the magnetic equilibrium, and returns plasma current, magnetic-axis coordinates, coil currents, the LCFS, and control-point fluxes. With a $1~\mathrm{ms}$ control period, the axis coordinates, coil currents, controlled current, and fluxes enter state estimation; LCFS flux supplies the isoflux reference. These are simulated counterparts of device signals, without execution of diagnostics or PTEFIT. Section 3 describes their correspondence to the planned device interfaces.

The three coupled components realize the state and output relations in Eqs.~\eqref{eq:nonlinear_state}--\eqref{eq:nonlinear_output}. Closed-loop validation uses nonlinear FGE, whereas the controller uses its linearized model for prediction and state estimation. The original model has 538 states, many representing unmeasurable vessel currents. Section 3 therefore constructs a reduced model and observer that retain control-relevant responses; Section 4 uses the estimated states in constrained MPC, and Section 5 evaluates tracking and computational cost.

\subsection{Observer-Based MPC Framework}
\label{sec:control_framework}

\begin{figure}[ht]
\centering
\includegraphics[width=0.95\textwidth]{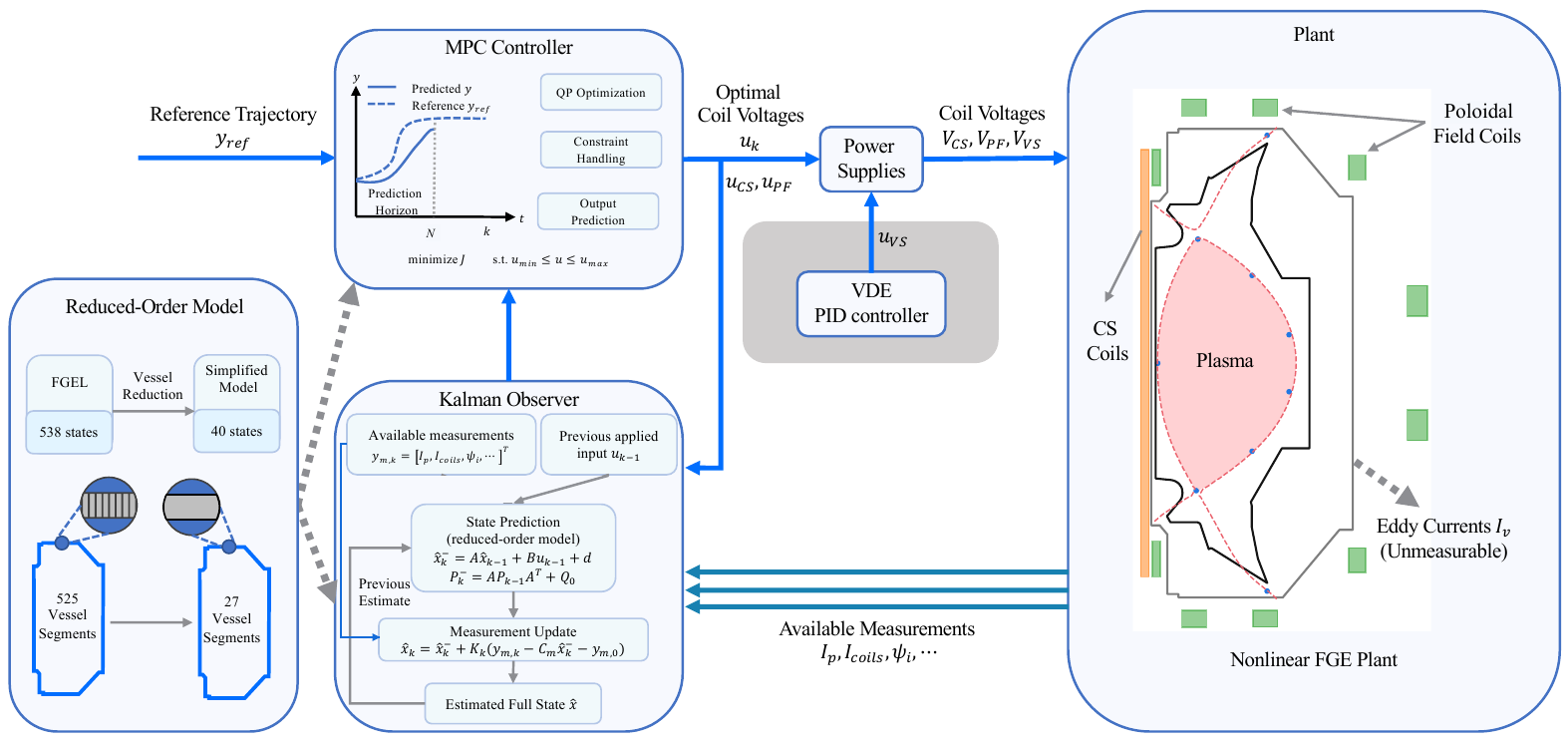}
\caption{Observer-based MPC architecture for isoflux control. The grey-shaded VS loop provides system context and lies outside the scope of this study.}
\label{fig:framework}
\end{figure}

Figure~\ref{fig:framework} connects the control task, prediction model, and feedback signals in an output-feedback MPC loop. The reduced model predicts current and flux responses, while the Kalman observer combines model information with available measurements to estimate all reduced-model states. MPC initializes its predictions from this estimate, solves the joint current and flux tracking problem under voltage constraints, and applies the current CS/PF voltage commands to the nonlinear FGE plant.

\FloatBarrier

Table~\ref{tab:nomenclature} lists the main notation. Parameter values are given with the corresponding method and simulation settings.
\begin{table}[htbp]
\centering
\caption{Main notation.}
\label{tab:nomenclature}
\small
\renewcommand{\arraystretch}{1.2}
\setlength{\tabcolsep}{4pt}
\begin{tabularx}{\textwidth}{@{}>{\raggedright\arraybackslash}p{2.1cm} >{\raggedright\arraybackslash}X@{\hspace{6mm}}>{\raggedright\arraybackslash}p{2.1cm} >{\raggedright\arraybackslash}X@{}}
\toprule
Symbol & Definition & Symbol & Definition \\
\midrule
\multicolumn{2}{@{}l}{\textbf{Plant and control targets}} & \multicolumn{2}{l@{}}{\textbf{Model and algorithm}}\\
$\Gamma_{\rm ref}$ & Desired plasma boundary & $k,T_s$ & Discrete time index, control period \\
$\mathbf{p}_i$ & Boundary control point $i$ & $n_x,N$ & State dimension, prediction steps \\
$N_c$ & Number of flux control points & $\mathbf{x}$ & Model state vector \\
$\psi(R,Z,t)$ & Poloidal flux field (Wb) & $f(\cdot),h(\cdot)$ & State-derivative and output functions \\
$\psi_i$ & Flux at control point $i$ (Wb) & $A_c,B_c;\ A,B$ & Continuous/discrete state and input matrices \\
$\psi_{\rm ref}$ & Common LCFS-flux reference (Wb) & $C,D$ & Output and feedthrough matrices \\
$\boldsymbol{\psi}_c$ & Control-point flux vector & $\mathbf{d},\mathbf{y}_0$ & Combined state offset, operating-point output \\
$e_i$ & Pointwise flux error (Wb) & $\mathbf{y}_m$ & Observer measurement vector \\
$\boldsymbol{e}_{\psi}$ & Control-point flux-error vector & $C_m,\mathbf{y}_{m,0}$ & Measurement matrix and offset \\
$I_p,I_p^{\rm ref}$ & Plasma current and reference (A) & $\hat{\mathbf{x}}_k^{-},\hat{\mathbf{x}}_k$ & Prior and posterior state estimates \\
$\mathbf{y}$ & Current and flux controlled outputs & $P_k^{-},P_k$ & Prior and posterior error covariances \\
$\mathbf{y}_{\rm ref}$ & Controlled-output reference vector & $Q_o,R_o$ & Process and measurement noise covariances \\
$\boldsymbol{e}_y$ & Joint current/flux tracking error & $\mathbf{r}_k,K_k$ & Innovation, Kalman gain \\
$\mathbf{u}$ & Selected nine-channel CS/PF voltage vector (V) & $Q,Q_f,R$ & Stage/terminal tracking and input weights \\
$\mathbf{u}_{\min},\mathbf{u}_{\max}$ & Single-step voltage bounds (V) & $\mathbf{X},\mathbf{Y},\mathbf{U}$ & Stacked predicted states, outputs, and inputs \\
$R_{\rm axis},Z_{\rm axis}$ & Magnetic-axis coordinates (m) & $\mathbf{F},\boldsymbol{\Phi}$ & State/input maps for output prediction \\
$I_{CS}$ & CS-coil current (A) & $\mathbf{Y}_0$ & Stacked output-prediction offset \\
$\mathbf{I}_{\mathrm{PF}}$ & PF-coil current vector (A) & $H,g$ & QP Hessian and linear term \\
$I_{VS}$ & VS-coil current (A) & $\mathbf{U}_{\min},\mathbf{U}_{\max}$ & Stacked input bounds \\
\bottomrule
\end{tabularx}
\end{table}

\section{Model Linearization, Reduction, and State Observation}

For the joint current and isoflux regulation task defined in Section 2, MPC solves a voltage-constrained optimal-control problem over a receding horizon. Online implementation requires a prediction model suitable for optimization and a current state from which to initialize predictions. This section develops the prediction model through FGE linearization, reduces its dimension through vacuum-vessel spatial coarsening to accommodate a millisecond computation budget, and constructs a Kalman observer for electromagnetic states that cannot be measured directly.

\subsection{FGE Linearization and Prediction Model}

MPC must predict plasma-current and boundary-flux responses to coil voltages. To incorporate these predictions into an online quadratic program, a linear state-space model is constructed near the target equilibrium, starting from the nonlinear FGE model\cite{FGE} in Eqs.~\eqref{eq:nonlinear_state}--\eqref{eq:nonlinear_output} of Section 2.2.

The target diverted equilibrium is chosen as the linearization point $(\mathbf{x}_L,\mathbf{u}_L)$. Local dynamics are expressed using the deviations $\Delta\mathbf{x}=\mathbf{x}-\mathbf{x}_L$ and $\Delta\mathbf{u}=\mathbf{u}-\mathbf{u}_L$, yielding
\begin{equation}
\Delta\dot{\mathbf{x}}=A_c\Delta\mathbf{x}+B_c\Delta\mathbf{u},
\label{eq:linearized_state}
\end{equation}
\begin{equation}
\Delta\mathbf{y}=C\Delta\mathbf{x}+D\Delta\mathbf{u},
\label{eq:linearized_output}
\end{equation}
where $\Delta\mathbf{y}$ is the deviation from the operating-point output, $A_c$ and $B_c$ are the continuous-time state and input matrices, and $C$ and $D$ are the output and feedthrough matrices. The FGE model used here has $D=0$, so the subsequent observer and MPC output equations contain no direct input term.

Equilibrium-sensitivity calculations yield the linearized FGE (FGEL) model. Its 538 states comprise plasma-equilibrium-related states, external coil-current states, and vessel-current states, describing the coupled electromagnetic responses of the plasma, external circuits, and conducting structures.

The discrete-time representation used for controller design is
\begin{equation}
\mathbf{x}_{k+1}=\mathbf{x}_0+A(\mathbf{x}_k-\mathbf{x}_{L})+B(\mathbf{u}_k-\mathbf{u}_{L})+\dot{\mathbf{x}}_{L},
\label{eq:discrete_state}
\end{equation}
\begin{equation}
\mathbf{y}_{k}=C(\mathbf{x}_k-\mathbf{x}_{L})+\mathbf{y}_{0}.
\label{eq:discrete_output}
\end{equation}
where $\mathbf{x}_{L}$ and $\mathbf{u}_{L}$ define the linearization point, $\mathbf{y}_{0}$ is the operating-point output, and $\mathbf{x}_0$ is a fixed baseline vector in the model's discrete-state representation, rather than the initial state of each receding-horizon prediction. Equation~\eqref{eq:discrete_state} retains the model-specific notation $\dot{\mathbf{x}}_{L}$ for a discrete-state offset; it does not denote a time derivative here. The matrices $A,B$ in this equation are discrete-time matrices. To obtain the affine form used by the observer and MPC, terms independent of the current state and input are combined as
\begin{equation}
\mathbf{d}=\mathbf{x}_0-A\mathbf{x}_L-B\mathbf{u}_L+\dot{\mathbf{x}}_L,
\label{eq:affine_state_bias}
\end{equation}
so that the state update becomes $\mathbf{x}_{k+1}=A\mathbf{x}_k+B\mathbf{u}_k+\mathbf{d}$. The output is still calculated from the state deviation relative to the operating point according to Eq.~\eqref{eq:discrete_output}.

\subsection{Vacuum-Vessel Spatial Coarsening and Model Reduction}

Linearization provides predictions suitable for optimization, but a high state dimension still increases the cost of state propagation, observer calculations, and prediction updates. For a $1~\mathrm{ms}$ control period, reduction must decrease the number of states involved in online calculations while retaining the response from coil voltages to plasma current and boundary flux. The vessel discretization is therefore coarsened, and its accuracy is assessed by comparing full-order and reduced-order input--output responses.

\subsubsection{Construction of the Reduced-Order Model}

To identify the subsystem to be reduced, the 538-state linearized FGE model is partitioned according to physical meaning as
\begin{equation}
\mathbf{x}=[\mathbf{x}_{c}^{T},\mathbf{x}_{coil}^{T},\mathbf{x}_{v}^{T}]^T,
\label{eq:state_classification}
\end{equation}
where $\mathbf{x}_{c}$ contains plasma-equilibrium-related states, $\mathbf{x}_{coil}$ contains active-coil current states, and $\mathbf{x}_{v}$ contains vessel-current states.

Of these states, 525 arise from the fine spatial discretization of the vessel. Because induced vessel currents contribute to transient electromagnetic responses, their influence on the controlled outputs must be retained. Since this study concerns plasma-current and boundary-flux regulation, model accuracy is assessed through control-relevant input--output responses rather than reconstruction of every local vessel-current component.

The EXL-50U vacuum vessel consists of 27 interconnected macroscopic segments, each further subdivided into smaller elements in the full-order FGE model. Here, each macroscopic segment is represented by a single element. The electromagnetic model is regenerated on the coarsened geometry and linearized near the target diverted equilibrium. This procedure retains the overall vessel geometry and connectivity, together with the states and physical descriptions of the plasma and active coils; the electromagnetic coupling matrices are recalculated from the coarsened model. This reduction of the vessel's spatial discretization resolution is referred to below as model reduction through vacuum-vessel spatial coarsening.

Coarsening reduces the number of vessel-current states from 525 to 27. The remaining 13 states consist of 1 plasma-related state and 12 current states for the CS, 10 PF coils, and VS coil, giving $n_x=40$. The model includes the VS current, whereas the MPC input comprises nine selected CS/PF voltage channels. These describe internal dynamics and adjustable actuators, respectively, and need not have the same dimension.

\begin{figure}[ht]
\centering
\includegraphics[width=0.95\textwidth, trim={2cm 8.5cm 2cm 6cm}, clip]{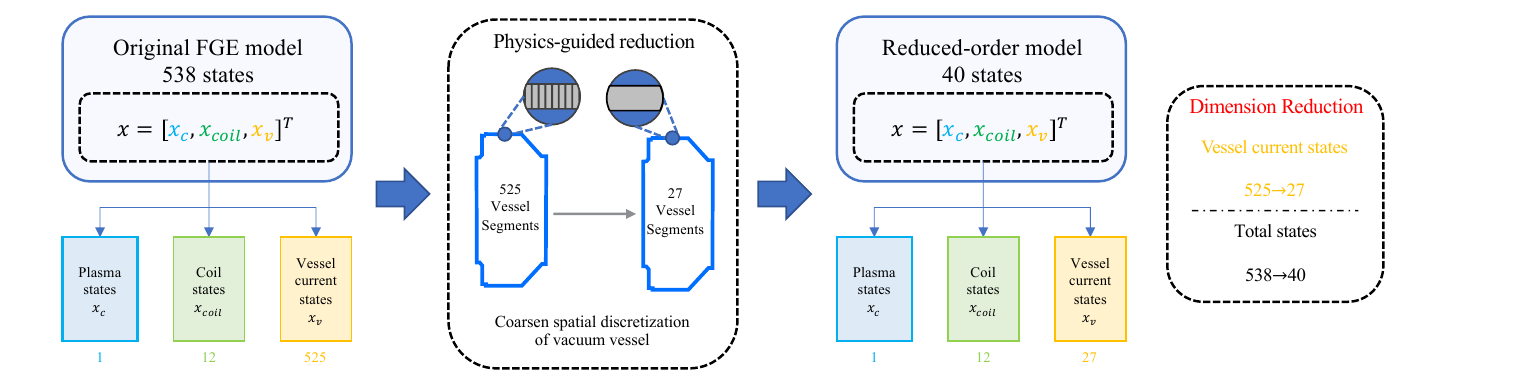}
\caption{FGE model reduction through vacuum-vessel spatial coarsening.}
\label{fig:ROM_reduction}
\end{figure}

Figure~\ref{fig:ROM_reduction} summarizes this construction. Reduction retains the vessel's contribution to the electromagnetic response while decreasing the degrees of freedom describing its spatial distribution. The effect on current and boundary-flux prediction is assessed below.

\subsubsection{Full-Order and Reduced-Order Response Validation}

To assess the effect of geometric coarsening on predicted responses, the 538-state full-order linear model (FOM) and 40-state reduced-order linear model (ROM) are compared. Both use consistent initial conditions and identical voltage inputs on the CS, PF, and VS channels. The complete coil-voltage vector used in this model-response test is denoted by
\[
\mathbf{u}_{\mathrm{test}}=
[u_{CS},u_{PF1},\ldots,u_{PF10},u_{VS}]^T.
\]
This test input includes the VS voltage to compare the electromagnetic responses of the two models. It differs from the CS/PF-only MPC input $\mathbf{u}$ in Eq.~\eqref{eq:input_vector} and is not the MPC decision vector. The comparison covers the complete input sequence and includes plasma current, representative boundary fluxes, and fluxes at all selected control points.

\begin{figure}[ht]
\centering
\includegraphics[width=0.95\textwidth]{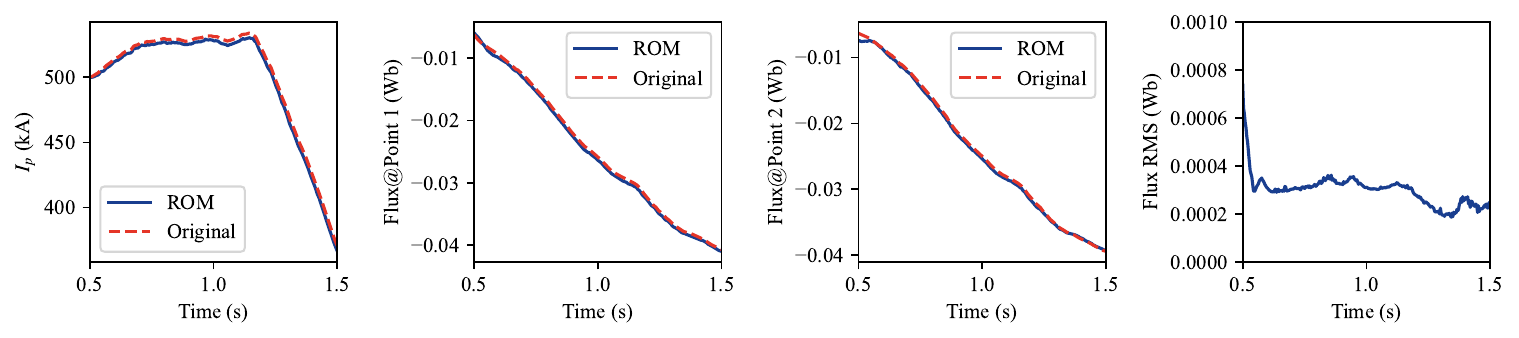}
\caption{Full-order (FOM) and reduced-order (ROM) model responses under consistent initial conditions and identical voltage inputs. From left to right: plasma current, fluxes at two representative control points, and the RMS difference in control-point fluxes.}
\label{fig:ROM_validation}
\end{figure}

Figure~\ref{fig:ROM_validation} shows the transient responses. The relative plasma-current and boundary-flux errors used to quantify trajectory differences are
\begin{equation}
E_{I_p}=\frac{\|I_p^{\mathrm{ROM}}-I_p^{\mathrm{FOM}}\|_2}
{\|I_p^{\mathrm{FOM}}\|_2},
\qquad
E_{\psi}=\frac{\|\boldsymbol{\psi}^{\mathrm{ROM}}-
\boldsymbol{\psi}^{\mathrm{FOM}}\|_F}
{\|\boldsymbol{\psi}^{\mathrm{FOM}}\|_F}.
\label{eq:rom_validation_metrics}
\end{equation}
where $I_p^{\mathrm{FOM}}$ and $I_p^{\mathrm{ROM}}$ are current sample sequences over the complete input sequence, and $\boldsymbol{\psi}^{\mathrm{FOM}}$ and $\boldsymbol{\psi}^{\mathrm{ROM}}$ are flux sample matrices for the same time instants and control points. These metrics use the actual current and flux outputs, not their deviations from the linearization point. The norms $\|\cdot\|_2$ and $\|\cdot\|_F$ denote the vector Euclidean norm and matrix Frobenius norm, respectively. Over $t=[0.5,1.5]~\mathrm{s}$, covering the complete input sequence, the calculated errors are $E_{I_p}=6.400395\times10^{-3}$ and
$E_{\psi}=1.047804\times10^{-2}$, or approximately $0.640\%$ and $1.048\%$, respectively.
Together with the plotted trajectories, these results show that the ROM closely reproduces the FOM current and boundary-flux responses over this test interval, supporting its use for equilibrium-control prediction in the present study.

This test assesses retention of the plasma-current and boundary-flux responses to CS, PF, and VS inputs. It does not reconstruct local vessel eddy-current distributions or separately validate fast vertical-stabilization performance. Reducing the model from 538 to 40 states decreases the scale of state propagation, observer calculations, and prediction updates; the computation time of the integrated framework is evaluated in Section 5.

The resulting 40-state model is constructed at the target diverted equilibrium and held fixed as the internal model of the observer and MPC. The comparison above assesses retention of the full-order linear responses. Closed-loop performance on the nonlinear plant and applicability away from the design equilibrium are evaluated in Section 5 using nonlinear FGE simulations and a limiter-to-divertor configuration-mismatch test.

\subsection{Kalman State Estimation}

The reduced model provides compact predictions, but each receding-horizon optimization still requires the current state as its initial condition. Coil currents are available from power-supply measurements, whereas the retained vessel-induced currents are not directly measurable. A Kalman observer therefore combines the 40-state model with available measurements to generate a full-state estimate including the unmeasurable electromagnetic components.

\subsubsection{Available Measurements and State-Estimation Interface}

The available measurement vector is defined as
\begin{equation}
\mathbf{y}_{m}=[I_p,R_{\rm axis},Z_{\rm axis},\psi_1,\cdots,
\psi_{N_c},I_{CS},\mathbf{I}_{\mathrm{PF}}^{T},I_{VS}]^T,
\label{eq:measurement_vector}
\end{equation}
where $\mathbf{I}_{\mathrm{PF}}=[I_{\mathrm{PF}1},\ldots,I_{\mathrm{PF}10}]^T$ contains the 10 PF-coil currents and $I_{VS}$ is the VS-coil current. The measurement vector consists of plasma current, magnetic-axis coordinates, nine control-point fluxes, and CS, PF, and VS currents. These current measurements belong to the observer feedback interface and are distinct from the MPC voltage decision variables; including the VS current does not alter the definition of the CS/PF control input.

In the planned device interface, magnetic-axis coordinates, LCFS flux, and control-point fluxes are supplied by the real-time PTEFIT equilibrium-reconstruction system\cite{ptefit}, while plasma current and coil currents come from diagnostic and power-supply measurements, respectively. The LCFS flux serves only as the common reference for the nine MPC flux channels and is not included in the observer measurement vector $\mathbf{y}_m$. Control references and state-estimation measurements thus have distinct roles.

In the present simulations, the nonlinear FGE plant supplies synthetic counterparts of these signals without directly executing PTEFIT. Magnetic-axis coordinates, control-point fluxes, and current measurements enter the observer, while the LCFS flux constructs the MPC reference. This mapping corresponds to the planned device interface, but the current tests do not include PTEFIT reconstruction errors or execution latency.

These measurements are used in the prediction and correction steps below. Their contribution to control performance is evaluated within the integrated observer--MPC closed loop.

\subsubsection{Kalman Prediction and Correction}

Using the combined offset in Eq.~\eqref{eq:affine_state_bias} for the reduced model, the observer's discrete-state equation is
\begin{equation}
\mathbf{x}_{k+1}=A\mathbf{x}_{k}+B\mathbf{u}_{k}+\mathbf{d},
\label{eq:observer_state}
\end{equation}
with the measurement model
\begin{equation}
\mathbf{y}_{m,k}=C_m\mathbf{x}_{k}+\mathbf{y}_{m,0},
\label{eq:observer_output}
\end{equation}
where $C_m$ maps model states to available measurements and $\mathbf{y}_{m,0}$ is the measurement offset. The Kalman prediction--correction procedure fuses model and measurement information\cite{Kalman1960,kalman}, with $Q_o$ and $R_o$ denoting the assumed process and measurement noise covariances. Under the assumed linear model and noise statistics, the Kalman update minimizes the estimation-error covariance.

The state and covariance prediction steps are
\begin{equation}
\hat{\mathbf{x}}_{k}^{-}=A\hat{\mathbf{x}}_{k-1}
+B\mathbf{u}_{k-1}+\mathbf{d},
\label{eq:kalman_predict}
\end{equation}
\begin{equation}
P_{k}^{-}=AP_{k-1}A^T+Q_o,
\label{eq:kalman_covariance}
\end{equation}
where $\hat{\mathbf{x}}_k^{-}$ and $P_k^{-}$ are the prior state estimate and its error covariance. Once the current measurement is available, the innovation, Kalman gain, and posterior estimate are calculated as
\begin{equation}
\mathbf{r}_{k}=\mathbf{y}_{m,k}-C_m\hat{\mathbf{x}}_{k}^{-}
-\mathbf{y}_{m,0},
\label{eq:kalman_innovation}
\end{equation}
\begin{equation}
K_{k}=P_{k}^{-}C_m^T(C_mP_{k}^{-}C_m^T+R_o)^{-1},
\label{eq:kalman_gain}
\end{equation}
\begin{equation}
\hat{\mathbf{x}}_{k}=\hat{\mathbf{x}}_{k}^{-}+K_{k}\mathbf{r}_{k}.
\label{eq:kalman_update}
\end{equation}
The posterior error covariance is updated as
\begin{equation}
P_k=(I-K_kC_m)P_k^{-}.
\label{eq:kalman_posterior_covariance}
\end{equation}
The corrected estimate $\hat{\mathbf{x}}_k$ includes all states of the reduced model, with unmeasurable components updated through model propagation and measurement innovations. Section 4 uses this estimate as the prediction initial condition and the compact linear model developed here to formulate MPC under coil-voltage constraints.

\section{Constrained Output-Feedback MPC}

Section 3 provides MPC with a compact prediction model and a current state estimate through linearization, vacuum-vessel spatial coarsening, and state observation. This section formulates the current and isoflux targets defined in Section 2 as a finite-horizon optimization problem with coil-voltage constraints, derives its quadratic-programming implementation, and describes the closed-loop connection to the observer.

The joint objectives in Section 2 are expressed using the controlled output and reference vectors
\begin{equation}
\mathbf{y}=[I_p,\psi_1,\ldots,\psi_{N_c}]^T,\qquad
\mathbf{y}_{\rm ref}=[I_p^{\rm ref},\mathbf{1}_{N_c}^T\psi_{\rm ref}]^T.
\label{eq:output_vector}
\end{equation}
Here, $\mathbf{1}_{N_c}$ is an $N_c$-dimensional vector of ones. With $\boldsymbol{\psi}_c=[\psi_1,\ldots,\psi_{N_c}]^T$, the flux-error vector is $\boldsymbol{e}_{\psi}=\boldsymbol{\psi}_c-\mathbf{1}_{N_c}\psi_{\rm ref}$ and the joint error is $\boldsymbol{e}_y=\mathbf{y}-\mathbf{y}_{\rm ref}$.

Experimental operating requirements exclude PF3 and PF4 from feedback regulation in this study, so they are not included in the MPC optimization. The nine optimized voltage channels are CS, PF1, PF2, and PF5--PF10:
\begin{equation}
\mathbf{u}=[u_{CS},u_{PF1},u_{PF2},u_{PF5},\ldots,u_{PF10}]^T\in\mathbb{R}^{9}.
\label{eq:input_vector}
\end{equation}
This vector specifies the selected CS/PF optimization inputs. The VS control loop is not included in the MPC design; the VS-current state and its measurement interface are retained in the model.

\subsection{Finite-Horizon Optimization Using State Estimates}

At sampling instant $k$, MPC initializes its predictions with the posterior state estimate $\hat{\mathbf{x}}_k$ from the Kalman observer and optimizes a future sequence of CS/PF coil voltages. The reduced discrete-time prediction model is
\begin{equation}
\mathbf{x}_{k+1}=A\mathbf{x}_{k}+B\mathbf{u}_{k}+\mathbf{d},
\label{eq:mpc_prediction_state}
\end{equation}
\begin{equation}
\mathbf{y}_{k}=C(\mathbf{x}_{k}-\mathbf{x}_{L})+\mathbf{y}_{0},
\label{eq:mpc_prediction_output}
\end{equation}
where $\mathbf{d}$ combines the state offsets according to Eq.~\eqref{eq:affine_state_bias} in Section 3, $\mathbf{x}_L$ is the linearization-point state, and $\mathbf{y}_0$ is the corresponding operating-point output. At each sampling instant, $\mathbf{x}_{k}=\hat{\mathbf{x}}_k$ initializes the prediction.

For a prediction horizon of $N$ steps, the predicted states, outputs, and candidate control inputs calculated at the current instant are stacked as
\begin{equation}
\mathbf{X}=[\mathbf{x}_{k+1}^T,\mathbf{x}_{k+2}^T,\cdots,\mathbf{x}_{k+N}^T]^T,
\label{eq:prediction_state_vector}
\end{equation}
\begin{equation}
\mathbf{Y}=[\mathbf{y}_{k+1}^T,\mathbf{y}_{k+2}^T,\cdots,\mathbf{y}_{k+N}^T]^T,
\label{eq:prediction_output_vector}
\end{equation}
\begin{equation}
\mathbf{U}=[\mathbf{u}_{k}^T,\mathbf{u}_{k+1}^T,\cdots,\mathbf{u}_{k+N-1}^T]^T.
\label{eq:prediction_input_vector}
\end{equation}
Future states and outputs are predictions based on the current state estimate and a candidate input sequence, not future measurements. Section 2 defines the current control target using fixed control points, a plasma-current reference, and the current LCFS flux. Here, the current output and this target are converted into the finite-horizon reference sequence required by the optimizer. The nine flux channels use an exponential transition from their current flux values toward the common target value. This generates numerical references over the prediction horizon without changing the spatial control points or using future LCFS measurements. Below, $\mathbf{y}_{\mathrm{ref},k+i}$ denotes the step-$i$ reference output generated at the current control instant.

The objective penalizes plasma-current and control-point flux tracking errors together with coil-voltage control effort:
\begin{equation}
\min_{\mathbf{U}}J=
\sum_{i=1}^{N}(\mathbf{y}_{k+i}-\mathbf{y}_{\mathrm{ref},k+i})^T
Q_i(\mathbf{y}_{k+i}-\mathbf{y}_{\mathrm{ref},k+i})
+\sum_{i=0}^{N-1}\mathbf{u}_{k+i}^TR\mathbf{u}_{k+i},
\label{eq:mpc_cost}
\end{equation}
where $Q_i=Q$ weights output tracking over the first $N-1$ steps, $Q_N=Q_f$ is the terminal tracking weight, and $R$ weights control effort. The matrices $Q$ and $Q_f$ balance the plasma-current and control-point flux tracking objectives, while $R$ balances tracking performance against voltage usage. The voltage penalty does not replace hard actuator constraints. The finite-horizon optimization also satisfies
\begin{equation}
\begin{aligned}
\mathbf{x}_{k}&=\hat{\mathbf{x}}_k,\\
\mathbf{x}_{k+i+1}&=A\mathbf{x}_{k+i}+B\mathbf{u}_{k+i}+\mathbf{d},\\
\mathbf{y}_{k+i+1}&=C(\mathbf{x}_{k+i+1}-\mathbf{x}_L)+\mathbf{y}_0,\\
\mathbf{u}_{\min}&\leq\mathbf{u}_{k+i}\leq\mathbf{u}_{\max},
\qquad i=0,\ldots,N-1.
\end{aligned}
\label{eq:mpc_horizon_constraints}
\end{equation}
These constraints incorporate the current state estimate, prediction dynamics, and CS/PF actuator limits into the same optimization problem.

Eliminating the predicted states gives the output trajectory
\begin{equation}
\mathbf{Y}=\mathbf{F}\hat{\mathbf{x}}_{k}
+\boldsymbol{\Phi}\mathbf{U}+\mathbf{Y}_{0},
\label{eq:prediction_matrix}
\end{equation}
where $\mathbf{F}$ maps the current state to future outputs and $\boldsymbol{\Phi}$ maps future inputs to outputs. With block indices $i,j=1,\ldots,N$, the corresponding blocks and offset terms are
\begin{equation}
\begin{aligned}
\mathbf{F}_i&=CA^i,\\
\boldsymbol{\Phi}_{ij}&=
\begin{cases}
CA^{i-j}B,&j\leq i,\\
0,&j>i,
\end{cases}\\
(\mathbf{Y}_0)_i&=C\sum_{\ell=0}^{i-1}A^\ell\mathbf{d}
+\mathbf{y}_0-C\mathbf{x}_L.
\end{aligned}
\label{eq:prediction_blocks}
\end{equation}
Thus, $\mathbf{Y}_{0}$ includes both the propagated state offset and the constant output term.

Stack the reference trajectory as $\mathbf{Y}_{\mathrm{ref}}=[\mathbf{y}_{\mathrm{ref},k+1}^T,\ldots,\mathbf{y}_{\mathrm{ref},k+N}^T]^T$ and define
\[
\bar{Q}=\operatorname{diag}(Q,\ldots,Q,Q_f),\qquad
\bar{R}=\operatorname{diag}(R,\ldots,R),
\]
with $N$ diagonal blocks in each matrix. Equation~\eqref{eq:mpc_cost} becomes
\begin{equation}
J=(\mathbf{Y}-\mathbf{Y}_{\mathrm{ref}})^T\bar{Q}
(\mathbf{Y}-\mathbf{Y}_{\mathrm{ref}})+\mathbf{U}^T\bar{R}\mathbf{U}.
\label{eq:mpc_stacked_cost}
\end{equation}
Substituting Eq.~\eqref{eq:prediction_matrix} and dropping terms independent of $\mathbf{U}$ yields the quadratic programming (QP) problem
\begin{equation}
\min_{\mathbf{U}}\frac{1}{2}\mathbf{U}^{T}H\mathbf{U}
+\mathbf{U}^{T}g,
\label{eq:qp_cost}
\end{equation}
subject to the CS/PF voltage constraints
\begin{equation}
\mathbf{U}_{\min}\leq\mathbf{U}\leq\mathbf{U}_{\max}.
\label{eq:qp_constraints}
\end{equation}
where $\mathbf{U}_{\min}$ and $\mathbf{U}_{\max}$ stack the single-step voltage bounds over the prediction horizon. The Hessian $H$ and linear term $g$ are
\begin{equation}
\begin{aligned}
H&=2(\boldsymbol{\Phi}^T\bar{Q}\boldsymbol{\Phi}+\bar{R}),\\
g&=2\boldsymbol{\Phi}^T\bar{Q}
(\mathbf{F}\hat{\mathbf{x}}_k+\mathbf{Y}_0-\mathbf{Y}_{\mathrm{ref}}).
\end{aligned}
\label{eq:qp_coefficients}
\end{equation}
The linear prediction model makes the output error an affine function of the input sequence, so the tracking objective becomes quadratic in $\mathbf{U}$ while the voltage limits remain linear inequalities. Joint current and isoflux regulation is therefore implemented as a QP with box constraints on the inputs.

\subsection{Online Implementation of the Observer and MPC}

\begin{figure}[ht]
\centering
\includegraphics[width=0.95\textwidth]{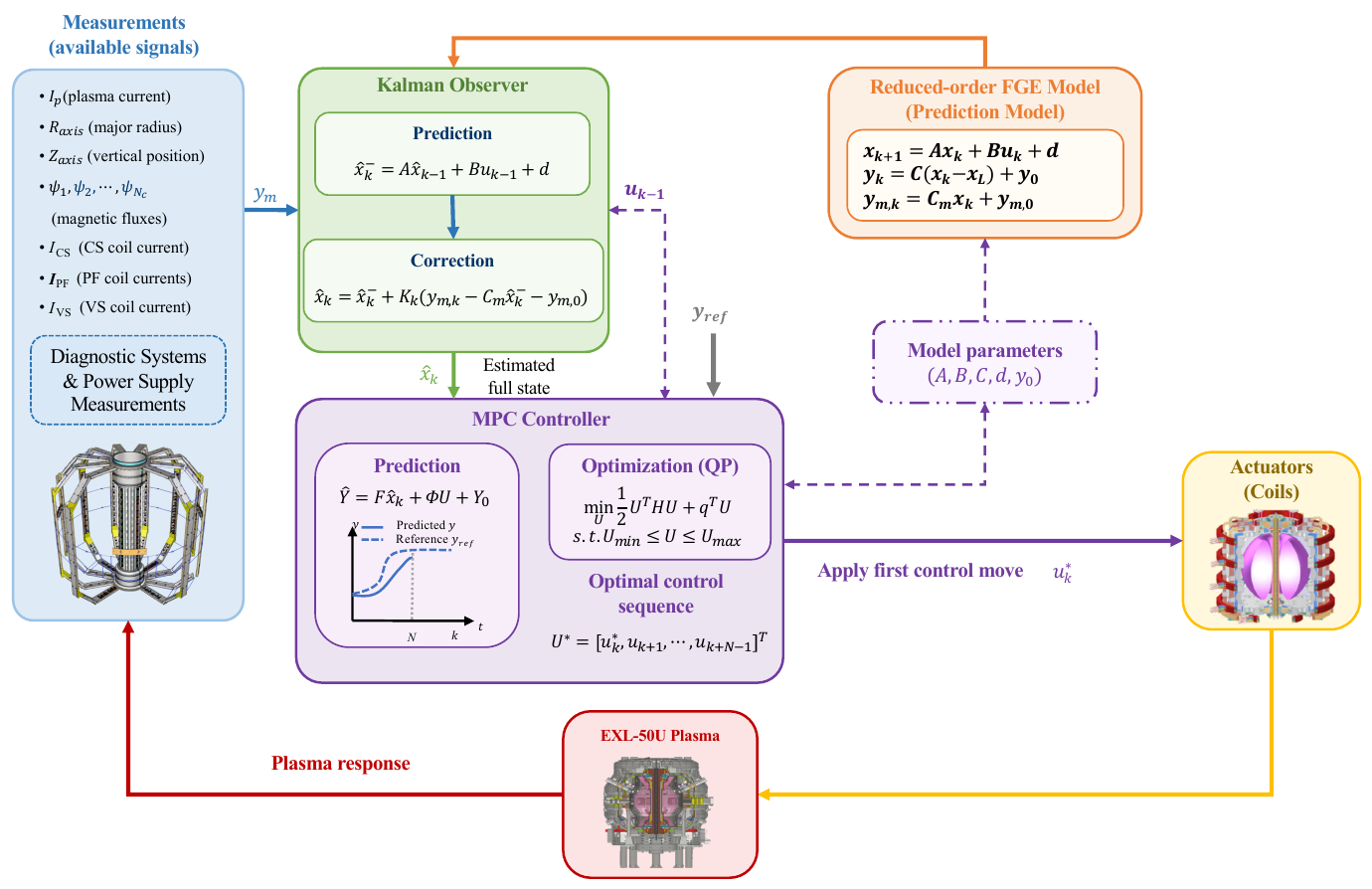}
\caption{Computational workflow of Kalman state estimation and constrained MPC. Device and diagnostic illustrations represent intended deployment interfaces; the present closed loop uses nonlinear FGE and its simulated feedback.}
\label{fig:adva_pipeline}
\end{figure}

Figure~\ref{fig:adva_pipeline} shows the output-feedback framework for real-time implementation. The Kalman observer combines available measurements with the reduced model to estimate all reduced-model states and initialize MPC prediction. Starting from this estimate, MPC jointly optimizes current and boundary-flux tracking over a finite horizon under CS/PF voltage constraints.

The OSQP solver\cite{OSQP} is selected for its warm-start capability to support millisecond-scale observation and optimization. Receding-horizon control repeatedly solves QPs with the same structure. Warm starting initializes the current iteration with the previous cycle's primal and dual solutions, reducing the computational burden of repeated solves; this is the main reason for selecting OSQP.

Matrix products, offset contributions, and the Hessian associated with the fixed prediction model, horizon, and weights are precomputed during initialization and reused. Online computations update only the linear term $g$ (denoted $q$ in the OSQP interface) using the current state estimate, measurements, and reference targets; fixed constraint bounds are not resubmitted at every step. A problem-specific OSQP C solver is generated and compiled with \texttt{-O3 -march=native} to reduce general-purpose interface overhead and exploit native instruction optimizations. Since the original OSQP backend is already implemented in C, the acceleration comes from specialization and optimization of the computation path rather than a change of programming language alone.

With $N=15$ and nine voltage inputs per step, the condensed QP contains $9\times15=135$ decision variables. Voltage bounds are represented using the identity constraint matrix $I_{135}$, giving 135 two-sided constraint rows and a $135+135=270$-dimensional internal linear system for the OSQP direct method. The structure of its factors is exploited to split each iterative linear solve into 135 scalar updates and a 135-dimensional triangular back-substitution, with the latter performed by optimized BLAS routines. Adaptive parameter changes still trigger refactorization and cache updates, and their cost is included in the timing. MPC runs in a dedicated CPU-pinned process with single-threaded BLAS, while FGE uses other CPUs to reduce interference. The process and thread configuration is unchanged between the baseline and optimized implementations.

These implementation changes preserve the MPC optimization problem, including the horizon, weights, constraints, double precision, solver tolerances, and iteration limit. Model reduction primarily decreases the cost of state propagation, observer calculations, and prediction updates; matrix precomputation, specialized code generation, and structured linear solves reduce online optimization cost. Section 5 reports both MPC computation time and communication-inclusive round-trip time relative to the $1~\mathrm{ms}$ cycle budget.

The implementation follows a certainty-equivalent approach: the optimizer treats $\hat{\mathbf{x}}_k$ as the current state, and the observer error covariance does not enter Eq.~\eqref{eq:mpc_cost} directly. Section 5 evaluates tracking and off-nominal behavior of this output-feedback implementation.

\begin{minipage}{\linewidth}
Full-state observation and receding-horizon optimization are connected within each control cycle as follows:
\begin{enumerate}
\item Acquire the measurements defined in Section 3 and the signals needed to construct the control reference. In these simulations, the nonlinear FGE plant returns the signals, and the LCFS flux is used only to construct the reference.
\item Update the full model-state estimate $\hat{\mathbf{x}}_k$ through the Kalman prediction--correction steps in Eqs.~\eqref{eq:kalman_predict}--\eqref{eq:kalman_posterior_covariance}, providing the initial condition for the current optimization.
\item Use the precomputed matrices, $\hat{\mathbf{x}}_k$, and the reference trajectory to update the QP linear term. Solve Eqs.~\eqref{eq:qp_cost}--\eqref{eq:qp_constraints} with warm-started OSQP to obtain the optimal sequence $\mathbf{U}^{*}$.
\item Apply the CS/PF voltage vector $\mathbf{u}^{*}_k$ for the current control step to the nonlinear FGE plant, and repeat estimation and optimization using new measurements at the next sampling instant.
\end{enumerate}
\end{minipage}

This integrates full-state observation, constrained optimal control, and computation suitable for real-time implementation into a complete output-feedback framework. State observation supplies the internal information needed for prediction, MPC coordinates current and isoflux regulation within actuator limits, and model reduction and solver acceleration reduce the online computational burden. The next section evaluates tracking, off-nominal responses, and MPC online computation time through nonlinear FGE closed-loop simulations.

\section{Nonlinear FGE Closed-Loop Simulation Results}

Nonlinear FGE simulations evaluate the proposed framework through nominal current and isoflux regulation, comparison with PID and LQR, and tests with measurement noise, configuration mismatch, a one-step delay, and combined noise and delay. MPC computation and communication-inclusive round-trip times are then compared with the $1~\mathrm{ms}$ control-period budget. These tests assess tracking, robustness under the specified non-nominal conditions, and computational feasibility.

\subsection{Simulation Settings and Evaluation Metrics}
\subsubsection{Closed-Loop Settings and Controller Comparison}

All MPC response curves and control-error metrics in this section use the computationally optimized implementation described in Section 4.2. This applies to all five conditions: nominal operation, measurement noise, configuration mismatch, delay, and combined noise and delay. The pre-optimization timing records retained in Section 5.5 serve only as a computational baseline.

The closed-loop plant is nonlinear FGE, while MPC uses the 40-state reduced linear model from Section 3 and the predictions in Eqs.~\eqref{eq:mpc_prediction_state}--\eqref{eq:mpc_prediction_output}. The tests thus evaluate a linear predictive controller acting on a nonlinear plant.

The initial equilibrium is obtained from EXL-50U discharge data. The physical current target is $I_p^{\rm ref}=500~\mathrm{kA}$, and nine fixed spatial control points ($N_c=9$), selected following Ref.~\cite{ptefit}, define the desired boundary. Their locations remain fixed; no additional time-varying geometric reference trajectory is prescribed. FGE supplies magnetic-axis coordinates, LCFS flux, and control-point fluxes as simulated counterparts of equilibrium-reconstruction outputs, without running PTEFIT. The LCFS flux supplies the common reference for the nine points and does not enter the observer measurement vector.

The prediction horizon is $N=15$ and the control period is $T_s=1~\mathrm{ms}$. The constrained problem is solved using warm-started OSQP. Comparisons use common valid records after controller activation. Nominal, mismatch, and delay cases each use 500 samples over $t=0.500$--$0.999~\mathrm{s}$. In the noise case, PID records end at $0.955~\mathrm{s}$, so all three controllers are evaluated over the same 456 samples from $0.500$ to $0.955~\mathrm{s}$. Complete LQR and MPC curves extend to $0.999~\mathrm{s}$. The combined-case window is specified in Section 5.4.4.

The controlled output is
\begin{equation}
\mathbf{y}=[I_p,\psi_1,\psi_2,\cdots,\psi_9]^T,
\end{equation}
where $\psi_i$ is the poloidal flux at control point $i$. The MPC inputs are the nine CS/PF voltage channels in Eq.~\eqref{eq:input_vector}.

PID and LQR are simulation baselines tuned under these common conditions, rather than direct experimental comparisons with the deployed controllers discussed in Section 1. All controllers use the same nonlinear FGE plant, initial equilibrium, current target, boundary-control points, LCFS reference, and voltage limits. PID uses available tracking errors directly; LQR and MPC share the linearized response model and Kalman state estimates. MPC enforces voltage bounds explicitly in its QP, whereas PID and LQR outputs are saturated at the same limits.

PID gains, LQR weights, and MPC weights and horizon are tuned independently in nominal nonlinear FGE simulations to coordinate current, LCFS, and control-point flux errors within the common voltage limits. These settings remain fixed in the non-nominal tests. The comparison therefore assesses tuned tracking performance and adaptability under common operating conditions.

\subsubsection{Tracking-Error Metrics}

Performance is evaluated using plasma-current, LCFS, and control-point flux errors. Let $N_c$ be the number of control points and $[t_a,t_b]$ the common evaluation interval after controller activation, containing $K$ samples indexed by $k=1,\ldots,K$. Metrics are calculated from the physical responses before measurement-noise addition and include the startup transient. The current reference is the physical target and is unaffected by internal reference compensation. For target point $\mathbf{p}_i$ and actual LCFS $\Gamma_k$, the nearest distance defines the instantaneous boundary error:
\begin{equation}
d_i(k)=\min_{\mathbf{q}\in\Gamma_k}\|\mathbf{p}_i-\mathbf{q}\|_2,\qquad
e_{\rm LCFS}(k)=\sqrt{\frac{1}{N_c}\sum_{i=1}^{N_c}d_i^2(k)}.
\end{equation}
\begin{equation}
\varepsilon_{\rm LCFS}^{\rm RMS}=\sqrt{\frac{1}{K}\sum_{k=1}^{K}e_{\rm LCFS}^2(k)},\qquad
\varepsilon_{I_p}^{\rm RMS}=\sqrt{\frac{1}{K}\sum_{k=1}^{K}\left[I_p(k)-I_p^{\rm ref}(k)\right]^2}.
\end{equation}
Section 2 defines the control objective using total poloidal flux. The figures and tables use the normalized flux error from the implementation, $\widetilde e_{\psi,i}(k)=[\psi_i(k)-\psi_{\rm ref}(k)]/(2\pi)$, equivalent to $(F_i-F_B)/(2\pi)$ in the records, reported in mWb. Figure~\ref{fig:MPC_result} shows individual selected flux-error channels; the five controller-comparison figures show all nine channels separately for each controller. The instantaneous RMS across channels can also be defined as
\begin{equation}
e_{\psi}^{\rm ch\text{-}RMS}(k)=
\sqrt{\frac{1}{N_c}\sum_{i=1}^{N_c}\widetilde e_{\psi,i}^2(k)}.
\end{equation}
The scalar flux metric in Tables~\ref{tab:equilibrium_metrics} and~\ref{tab:noise_delay_metrics} is instead the largest temporal RMS among the $N_c$ channels:
\begin{equation}
\varepsilon_{\psi}^{\rm max\text{-}RMS}=\max_{i=1,\ldots,N_c}\sqrt{\frac{1}{K}\sum_{k=1}^{K}\widetilde e_{\psi,i}^2(k)}.
\end{equation}
The former summarizes channels at each instant, whereas the latter identifies the worst individual channel over the evaluation interval. The tables use the latter, not a joint RMS over time and channels. Each condition uses a common window for all three controllers, as specified in Sections 5.1.1 and 5.4.4.

\subsection{Nominal Current and Shape Tracking}

\begin{figure}[h]
\centering
\includegraphics[width=0.8\textwidth]{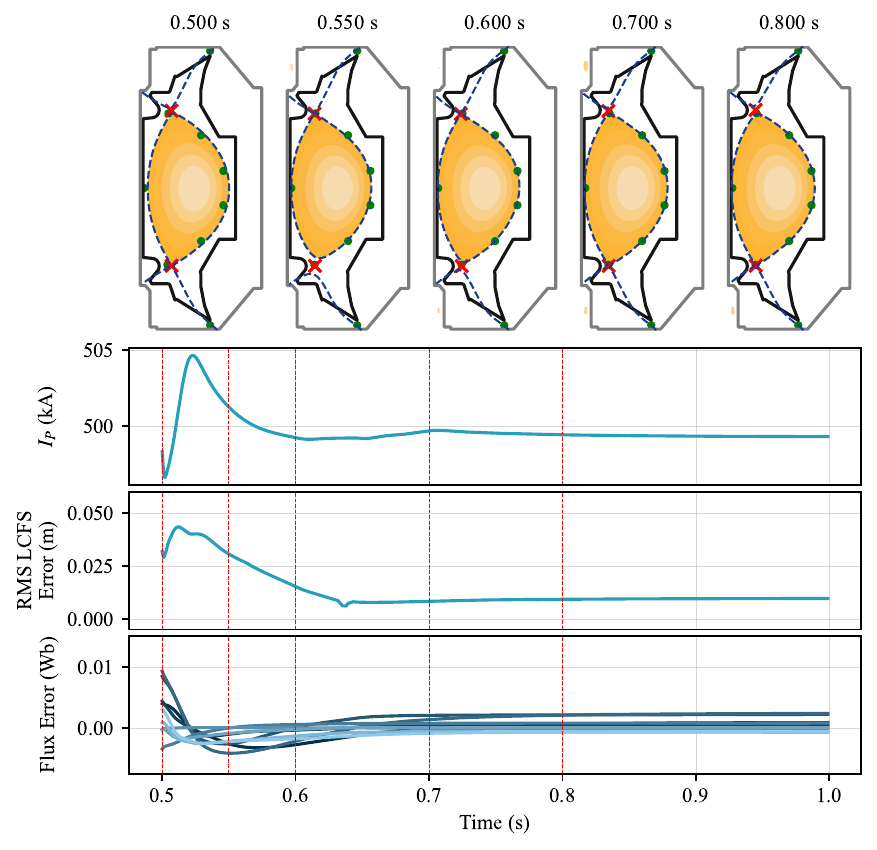}
\caption{Nominal MPC current and boundary tracking. From top to bottom: plasma-boundary evolution, plasma current, LCFS RMS error, and flux errors at selected isoflux control points.}
\label{fig:MPC_result}
\end{figure}

The nominal test examines simultaneous regulation of plasma current and boundary flux on the nonlinear plant. Figure~\ref{fig:MPC_result} shows the boundary evolution, current response, and tracking errors.

Following activation near $t=0.5~\mathrm{s}$, the boundary evolves from the initial equilibrium toward the target isoflux configuration. Plasma current initially overshoots, then decreases toward a quasi-steady value near $I_p^{\rm ref}=500~\mathrm{kA}$. The LCFS error decreases during the transition and approaches a low steady level. Control-point flux errors also decrease and remain bounded after the transient. These responses demonstrate joint current and nine-point isoflux regulation under the tested nominal conditions.

\subsection{Comparison with PID and LQR}

\begin{figure}[htbp]
\centering
\includegraphics[width=\textwidth]{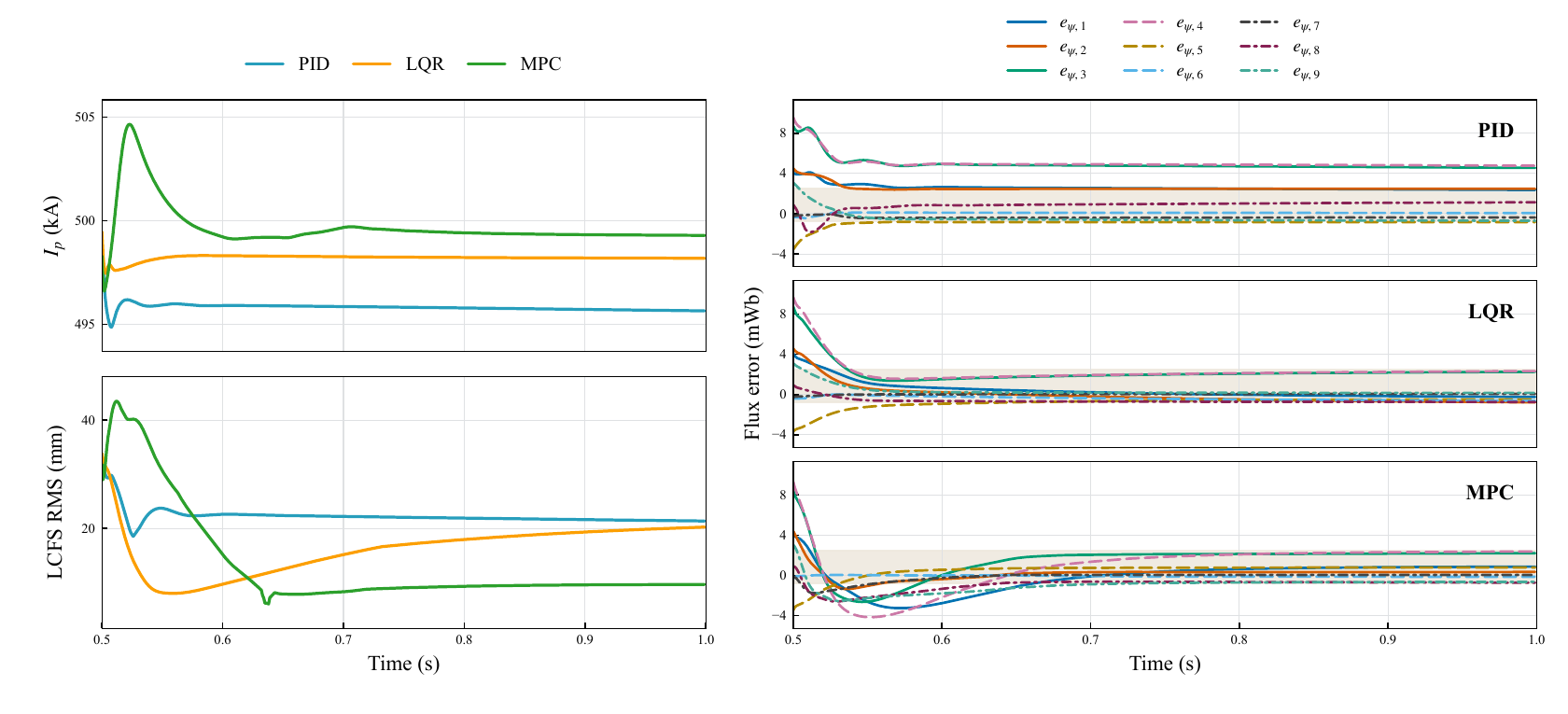}
\caption{Nominal PID, LQR, and MPC performance. Left: current and LCFS error. Right: nine flux-error channels for each controller. Shading denotes a visual reference band.}
\label{fig:controller_comparison}
\end{figure}

Figure~\ref{fig:controller_comparison} compares current and LCFS responses on the left and the nine flux-error channels for each controller on the right; Table~\ref{tab:equilibrium_metrics} provides the metrics. Shaded bands in the five comparison figures are visual references in mWb, not allowable-error bounds or acceptance criteria. Their ranges are $[-0.75,2.50]$, $[-1.00,3.00]$, $[-1.00,1.50]$, $[-1.00,2.00]$, and $[-1.00,2.00]$ for nominal, noise, mismatch, delay, and combined noise--delay conditions, respectively.

All three controllers remain bounded over the tested interval and drive the boundary toward the target configuration. PID provides feedback regulation but exhibits larger boundary and flux errors, while LQR produces a smooth transient response.

MPC achieves an LCFS RMS error of $0.0164~\mathrm{m}$, below PID ($0.0223~\mathrm{m}$) and slightly below LQR ($0.0168~\mathrm{m}$). Its current RMS error of $1.126~\mathrm{kA}$ and maximum-channel flux RMS error of $2.463~\mathrm{mWb}$ are also lower than both baselines. Under the selected tuning and constraints, MPC yields the lowest nominal error in all three metrics, supporting its use for joint current and boundary regulation in future EXL-50U experiments.

\subsection{Evaluation under Non-Nominal Conditions}

Measurement noise, cross-configuration model mismatch, and a one-step delay are introduced separately, followed by combined noise and delay. Controller parameters remain fixed after nominal tuning. Changes in tracking errors and the ability to sustain regulation assess robustness to the specified departures in measurement quality, initial configuration, and timing.

\subsubsection{Measurement Noise}

Independent Gaussian noise sequences are generated for plasma-current and flux measurements and normalized to specified maximum absolute amplitudes: $5~\mathrm{kA}$ for $I_p$ and $5\times10^{-4}~\mathrm{Wb}$ for each flux channel. No additional noise is applied to magnetic-axis or coil-current measurements. All three controllers use the same fixed noise realization. Noisy measurements enter feedback and Kalman estimation.

\begin{figure}[htbp]
\centering
\includegraphics[width=\textwidth]{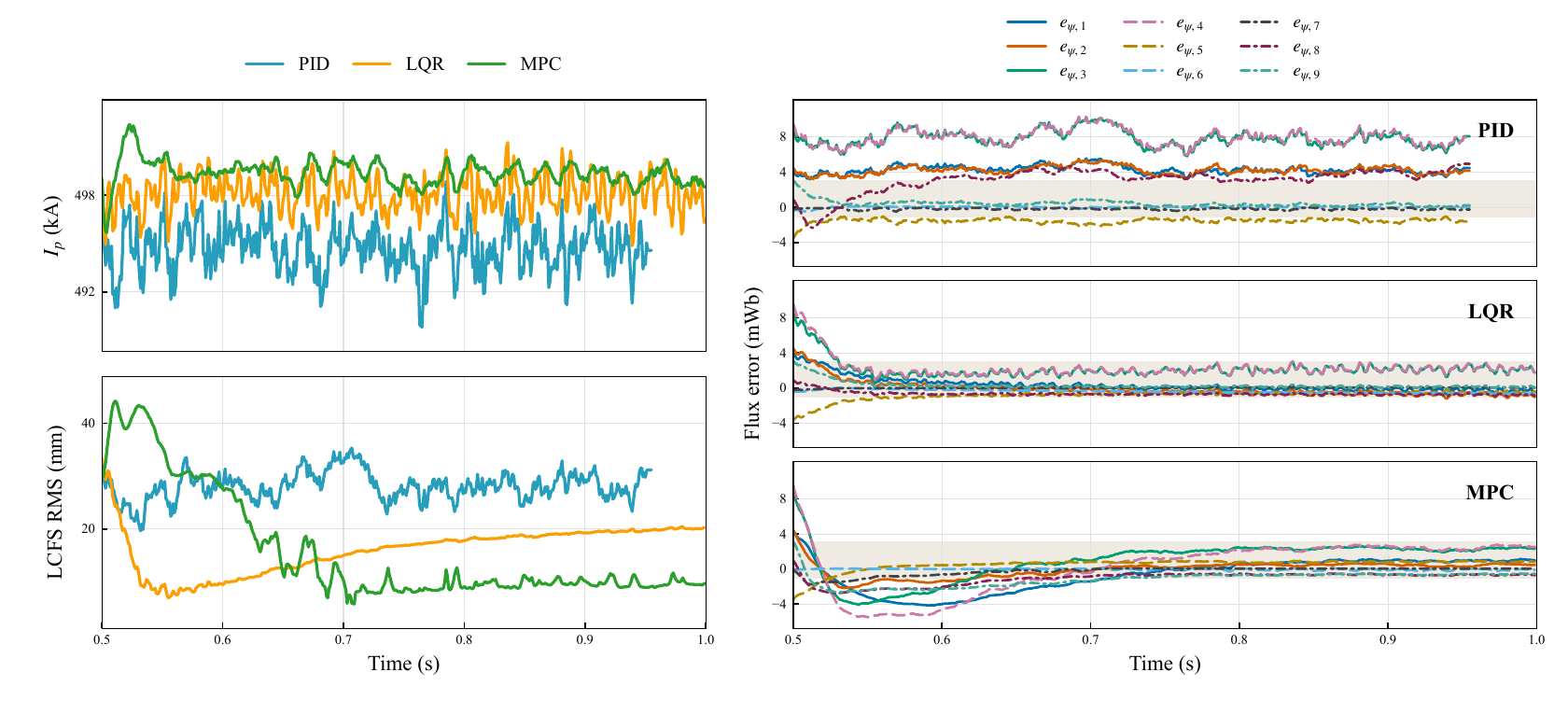}
\caption{PID, LQR, and MPC performance with measurement noise.}
\label{fig:measurement_noise}
\end{figure}

Figure~\ref{fig:measurement_noise} shows additional current and boundary fluctuations. In the later part of the PID run, a physically reasonable equilibrium solution can no longer be obtained and the plasma configuration cannot be maintained, causing early termination with the last valid record at $0.955~\mathrm{s}$. LQR and observer-based MPC continue to $0.999~\mathrm{s}$; MPC sustains equilibrium regulation with bounded tracking errors throughout the full test.

Within the common noisy window, MPC has the lowest current RMS error, $1.009~\mathrm{kA}$. Its LCFS and maximum-channel flux RMS errors are $0.0199~\mathrm{m}$ and $3.012~\mathrm{mWb}$, respectively, below PID and above LQR. MPC thus retains its current-tracking advantage while sustaining boundary and flux regulation. These closed-loop results support tolerance to the prescribed measurement noise and provide a simulation basis for future device tests.
\begin{table}[htbp]
\centering
\caption{Tracking errors under four conditions. Each condition uses a common evaluation window: 456 samples for measurement noise and 500 for the other conditions.}
\label{tab:equilibrium_metrics}
\small
\renewcommand{\arraystretch}{1.15}
\setlength{\tabcolsep}{7pt}
\begin{tabular}{llccc}
\toprule
Condition & Controller & $\varepsilon_{I_p}^{\rm RMS}$ (kA) & $\varepsilon_{\rm LCFS}^{\rm RMS}$ (m) & $\varepsilon_{\psi}^{\rm max\text{-}RMS}$ (mWb)\\
\midrule
Nominal & PID & 4.186 & 0.0223 & 5.090\\
 & LQR & 1.801 & 0.0168 & 2.578\\
 & MPC & 1.126 & 0.0164 & 2.463\\
\midrule
Noise & PID & 5.538 & 0.0282 & 8.024\\
 & LQR & 2.170 & 0.0163 & 2.600\\
 & MPC & 1.009 & 0.0199 & 3.012\\
\midrule
Mismatch & PID & 3.766 & 0.0264 & 5.572\\
 & LQR & 1.939 & 0.0244 & 1.886\\
 & MPC & 3.630 & 0.0267 & 2.935\\
\midrule
$1~\mathrm{ms}$ delay & PID & 4.193 & 0.0223 & 5.135\\
 & LQR & 1.830 & 0.0167 & 2.570\\
 & MPC & 1.049 & 0.0155 & 2.238\\
\bottomrule
\end{tabular}
\end{table}

\subsubsection{Model Mismatch across Plasma Configurations}

The reduced model, observer, and MPC designed for the target divertor equilibrium are applied to a nonlinear FGE plant initialized in a limiter configuration. The closed loop must complete the limiter-to-divertor transition without online relinearization, model reconstruction, or controller retuning. The difference between initial and model-design equilibria provides the mismatch.
\begin{figure}[htbp]
\centering
\includegraphics[width=\textwidth]{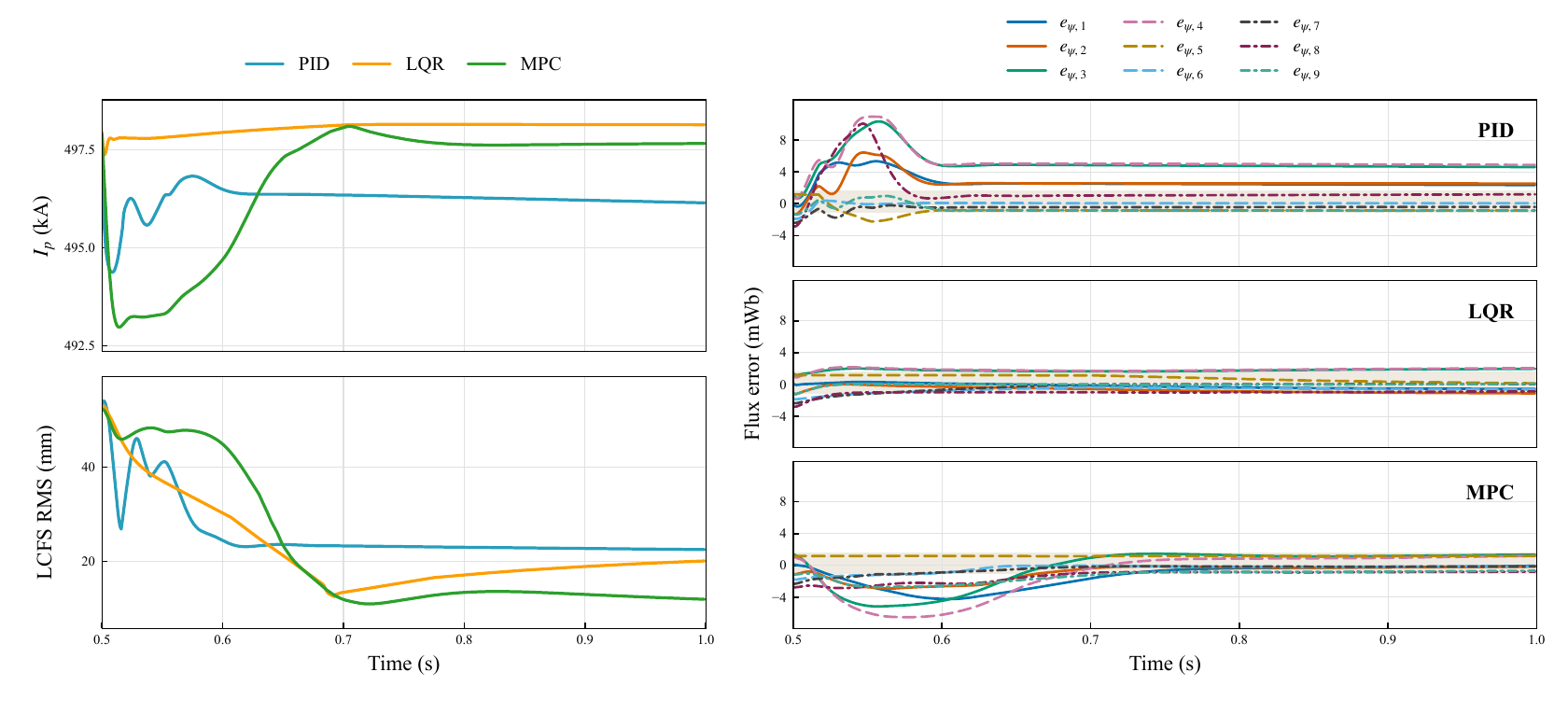}
\caption{PID, LQR, and MPC performance under configuration mismatch. Controllers retain divertor settings while the plant starts from a limiter configuration.}
\label{fig:model_mismatch}
\end{figure}

The mismatch produces larger transient deviations than nominal operation. Nevertheless, observer-based MPC remains bounded and progressively drives the plasma toward the target divertor configuration. LCFS and flux errors decrease as the plasma approaches the target equilibrium.

MPC current, LCFS, and maximum-channel flux RMS errors are $3.630~\mathrm{kA}$, $0.0267~\mathrm{m}$, and $2.935~\mathrm{mWb}$, respectively. Current and flux errors are below PID but above LQR; LCFS error is slightly above both baselines. Completing the transition with a fixed model and fixed parameters demonstrates applicability to the tested departure from the design equilibrium and provides a basis for subsequent EXL-50U validation.

\subsubsection{One-Step Delay}

The equilibrium controller receives state information delayed by one sampling period, $T_s=1~\mathrm{ms}$, to assess sensitivity to feedback timing. The VS loop is excluded from this delay test. PID, LQR, and MPC are compared using current, LCFS, and flux errors, with quantitative results in Table~\ref{tab:equilibrium_metrics}.
\begin{figure}[htbp]
\centering
\includegraphics[width=\textwidth]{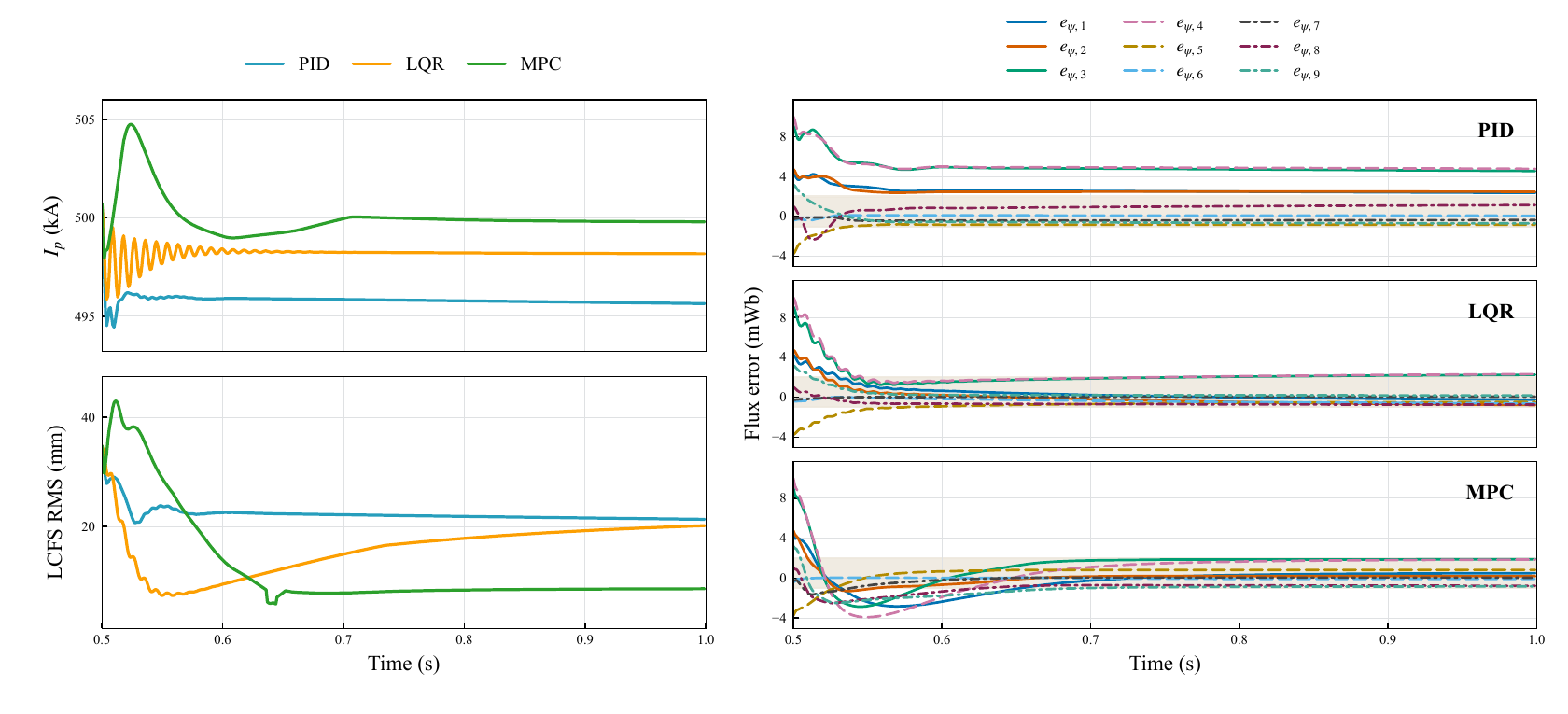}
\caption{PID, LQR, and MPC performance with a one-step delay ($1~\mathrm{ms}$).}
\label{fig:delay_test}
\end{figure}

The delay changes transient responses, but all three controllers remain bounded over the tested interval. MPC continues to drive current and boundary toward their targets. Its current, LCFS, and maximum-channel flux RMS errors are $1.049~\mathrm{kA}$, $0.0155~\mathrm{m}$, and $2.238~\mathrm{mWb}$, respectively, below both baselines and slightly below the corresponding nominal results. The sustained regulation and low errors support further timing design and experimental evaluation of the framework.

\subsubsection{Combined Measurement Noise and Delay}

\begin{figure}[htbp]
\centering
\includegraphics[width=\textwidth]{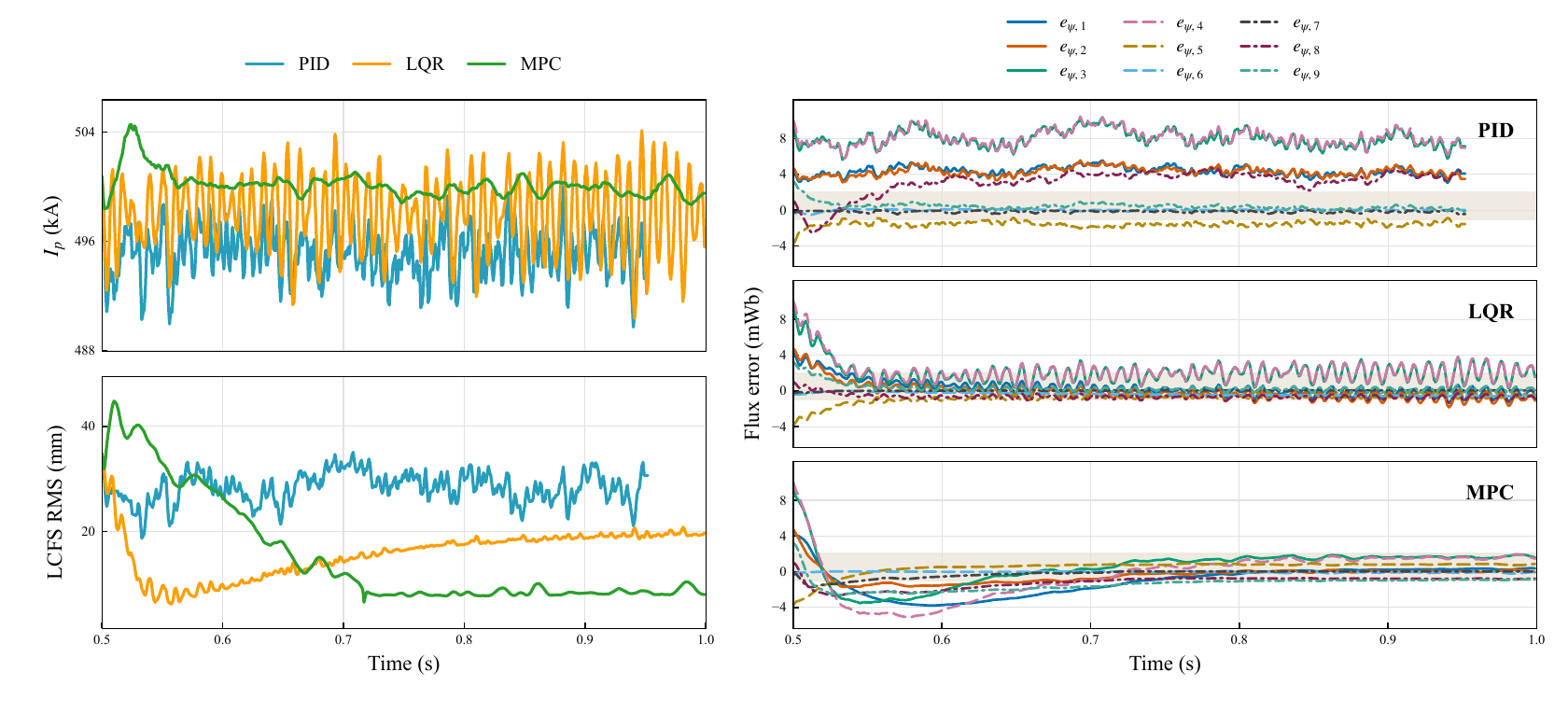}
\caption{Controller performance with combined measurement noise and a $1~\mathrm{ms}$ delay.}
\label{fig:noise_delay_test}
\end{figure}

The preceding measurement noise and $1~\mathrm{ms}$ delay are combined without configuration mismatch. The nominal configuration is retained, with random seed 42 and 500 planned control steps, as shown in Fig.~\ref{fig:noise_delay_test}.

The PID physical state stagnates after step 450, and the run exits after saving three repeated states. Excluding these records, all controllers are compared over the first 450 steps ($0.500$--$0.949~\mathrm{s}$). Complete LQR and MPC curves extend to $0.999~\mathrm{s}$. Table~\ref{tab:noise_delay_metrics} uses the metrics defined in Section 5.1.2.
\begin{table}[htbp]
\centering
\caption{Tracking errors with combined noise and delay over the common first 450 steps.}
\label{tab:noise_delay_metrics}
\small
\begin{tabular}{lccc}
\toprule
Controller & $I_p$ RMSE (kA) & LCFS RMS (m) & Max. channel flux RMSE (mWb)\\
\midrule
PID & 5.313 & 0.02850 & 8.156\\
LQR & 3.299 & 0.01607 & 2.631\\
MPC & 0.964 & 0.01936 & 2.699\\
\bottomrule
\end{tabular}
\end{table}

MPC achieves a current RMSE of $0.964~\mathrm{kA}$, lower than PID and LQR, and LCFS and flux errors below PID. The complete responses show sustained regulation under the combined disturbance, small current fluctuations after startup, and a low LCFS error in the later interval. These results support future joint current and boundary-control experiments on EXL-50U.

The nine-channel MPC errors over the last 100 steps range from $-0.9764$ to $1.9402~\mathrm{mWb}$. Rounding these limits outward gives the reference band $[-1.00,2.00]~\mathrm{mWb}$, with the same visual meaning as above.

\subsection{Online Computation Time and Real-Time Feasibility}

Computation times are recorded for 500 control steps in each of the five conditions. MPC computation time includes the current-step vector calculations, solver update, and QP solution, including refactorization and cache updates triggered by adaptive parameter changes. It excludes state observation, interprocess communication, and FGE simulation. Communication-inclusive round-trip times are recorded separately to assess control-process invocation overhead.

The prediction horizon, weights, and input constraints remain unchanged before and after optimization. Only the computational implementation is optimized; the earlier implementation is retained as the baseline.
\begin{table}[htbp]
\centering
\caption{MPC computation before and after implementation optimization (ms). Exceedances count samples above 1 ms; each group contains 500 steps. For combined noise and delay, the baseline is a rerun and the optimized results are saved records, with the same timing scope.}
\label{tab:computation}
\small
\renewcommand{\arraystretch}{1.15}
\setlength{\tabcolsep}{3pt}
\begin{tabular}{lcccccccc}
\toprule
& \multicolumn{4}{c}{Baseline MPC} & \multicolumn{4}{c}{Optimized MPC}\\
\cmidrule(lr){2-5}\cmidrule(lr){6-9}
Condition & Mean & P99 & Max. & \shortstack{$>1$ ms\\Count (\%)} & Mean & P99 & Max. & \shortstack{$>1$ ms\\Count (\%)}\\
\midrule
Nominal & 0.827 & 1.375 & 3.832 & 18 (3.6\%) & 0.422 & 0.570 & 2.344 & 1 (0.2\%)\\
Noise & 0.847 & 1.272 & 4.135 & 72 (14.4\%) & 0.441 & 0.700 & 2.642 & 2 (0.4\%)\\
Mismatch & 0.797 & 1.035 & 1.167 & 12 (2.4\%) & 0.413 & 0.534 & 0.636 & 0 (0.0\%)\\
$1~\mathrm{ms}$ delay & 0.785 & 1.207 & 3.841 & 16 (3.2\%) & 0.430 & 0.673 & 2.231 & 2 (0.4\%)\\
Noise + $1~\mathrm{ms}$ delay & 0.956 & 1.546 & 4.861 & 208 (41.6\%) & 0.427 & 0.666 & 2.779 & 2 (0.4\%)\\
\bottomrule
\end{tabular}
\end{table}

For the first four conditions, mean MPC computation time decreases by approximately 45.2\%--48.9\% relative to the baseline. Across all five cases, the optimized means are $0.413$--$0.441~\mathrm{ms}$ and P99 values are $0.534$--$0.700~\mathrm{ms}$. Counts above $1~\mathrm{ms}$ are 1, 2, 0, 2, and 2, corresponding to $0.2\%$, $0.4\%$, $0.0\%$, $0.4\%$, and $0.4\%$. Maximum times in the nominal, noise, and delay cases are $2.344$, $2.642$, and $2.231~\mathrm{ms}$, respectively, identifying the small number of tail events retained in the statistics.

In the combined case, mean, P99, and maximum times are $0.427$, $0.666$, and $2.779~\mathrm{ms}$. The two exceedances occur at steps 1 and 65, with times of $2.779$ and $1.268~\mathrm{ms}$ and iteration counts of 225 and 75. The baseline rerun gives a mean of $0.956~\mathrm{ms}$, P99 of $1.546~\mathrm{ms}$, maximum of $4.861~\mathrm{ms}$, and 208 exceedances ($41.6\%$). Relative to that rerun, the saved optimized records show a 55.3\% reduction in mean computation time and an exceedance rate of $0.4\%$.
\begin{figure}[htbp]
\centering
\includegraphics[width=0.90\textwidth]{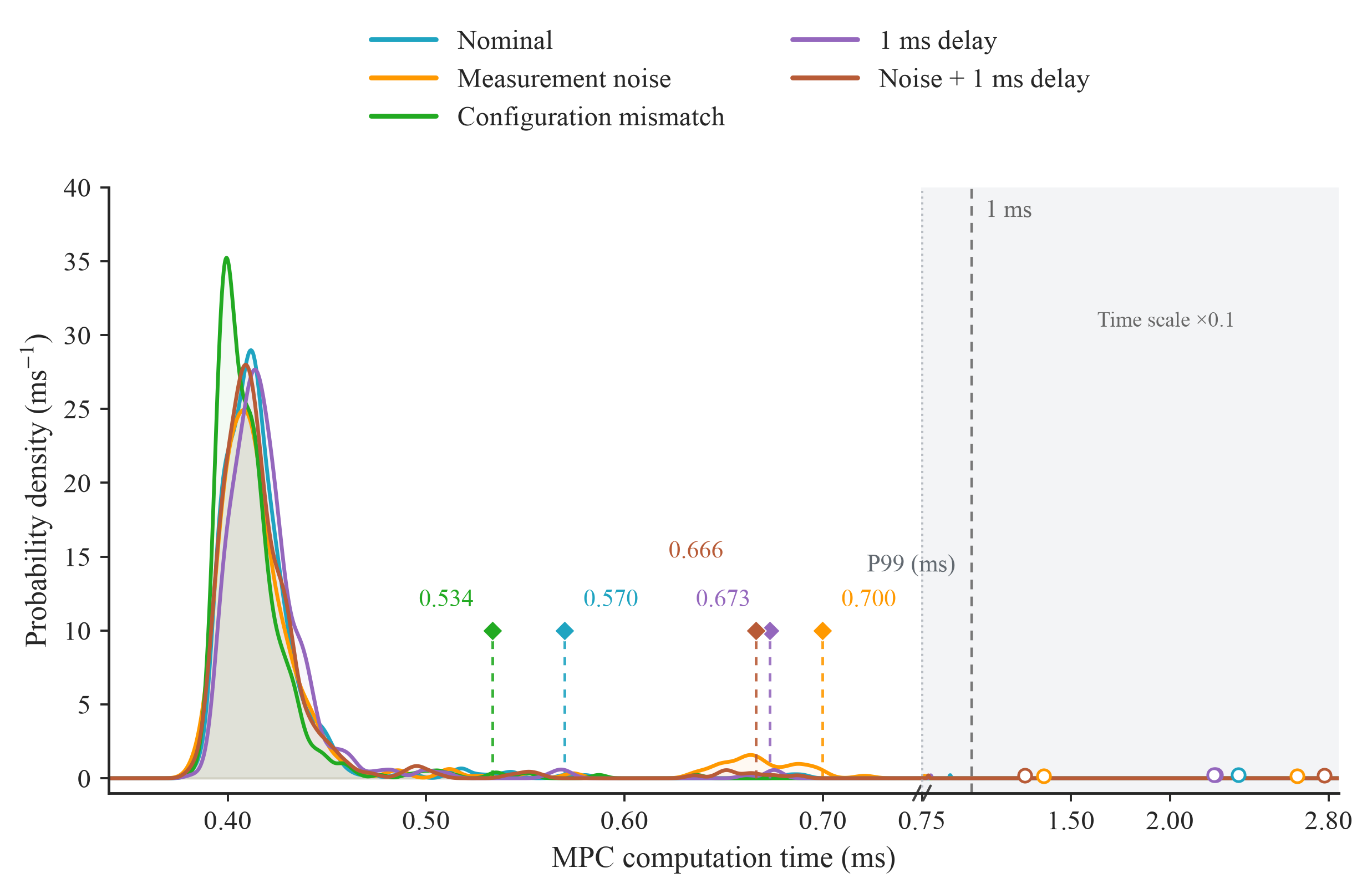}
\caption{Probability density of MPC computation time under five conditions.}
\label{fig:mpc_timing}
\end{figure}

Figure~\ref{fig:mpc_timing} uses the same stepwise records as the optimized columns of Table~\ref{tab:computation}. Gaussian kernel density estimates use all 500 samples per condition, including startup and outliers, with bandwidths chosen separately by the robust Silverman rule. The horizontal axis covers all measured times using a continuous piecewise-linear scale: above $0.75~\mathrm{ms}$, display width per unit time is one tenth of that below the breakpoint. No samples or time intervals are omitted, and density remains expressed in the original millisecond units. Colored diamonds mark P99 at a common density height of 10, with vertical dashed guides; their height does not represent the density at P99. Open circles show measured samples above $1~\mathrm{ms}$, and the vertical budget line marks the control period. The lower axis limit of $-1~\mathrm{ms}^{-1}$ provides space below the baseline; estimated densities remain nonnegative.

For the first four conditions, communication-inclusive round-trip P99 values are $1.070$, $1.202$, $0.959$, and $1.076~\mathrm{ms}$. The implementation substantially reduces online MPC computation and supports millisecond-scale control development, providing a basis for further platform-specific communication and execution scheduling and for EXL-50U closed-loop experiments.

\section{Conclusion}

An observer-based constrained MPC framework has been developed for isoflux control of the EXL-50U spherical tokamak. It combines plasma-current and nine-point boundary-flux regulation in a finite-horizon optimization problem with explicit CS/PF voltage bounds. Nonlinear FGE simulations demonstrate joint current and shape regulation. Under common nominal test conditions, the LCFS RMS error is $0.0164~\mathrm{m}$, below PID and slightly below LQR; current and maximum-channel flux RMS errors are also lower than both baselines. These results demonstrate the potential of MPC for coordinated boundary and current tracking under actuator constraints.

To support millisecond-scale control, vessel coarsening and electromagnetic model reconstruction reduce the linearized model from 538 to 40 states. Response comparisons verify retention of control-relevant current and flux dynamics. A Kalman observer supplies the full reduced-model state estimate for output-feedback MPC. Fixed-matrix precomputation, specialized OSQP code generation, and structured linear-system solution reduce online computation while preserving the control parameters. Mean computation time decreases by approximately 45.2\%--48.9\% for the first four conditions and by 55.3\% relative to the rerun baseline for combined noise and delay. Across all five conditions, mean times are $0.413$--$0.441~\mathrm{ms}$ and P99 values are $0.534$--$0.700~\mathrm{ms}$, covering current-step vector calculation, solver update, and solution. This computational performance supports millisecond-scale MPC execution and subsequent device deployment.

The non-nominal tests further demonstrate robustness of the complete observer--MPC loop under the specified conditions. With measurement noise, MPC achieves the lowest current error over the common window while sustaining boundary regulation. It completes the limiter-to-divertor transition without model relinearization or controller retuning. With a one-step delay, all three tracking errors remain below those of the baseline controllers. Under combined noise and delay, MPC also sustains current and boundary regulation. Together, these tests provide a simulation basis for future EXL-50U experiments involving measurement uncertainty, local model mismatch, and timing effects.

The resulting model reduction, state estimation, constrained optimization, and nonlinear validation provide both an algorithmic and a computational basis for on-device experiments. Future work will integrate the observer--MPC framework with PTEFIT real-time equilibrium reconstruction, diagnostics, and CS/PF power supplies under a common execution schedule. Joint current and boundary-control experiments on EXL-50U will then assess tracking and operational adaptability in actual discharges, progressing from simulation validation to device application.

\clearpage  % flush any remaining floats before the bibliography
\bibliographystyle{unsrt}
\bibliography{references}

\end{document}